\documentclass[aps,twocolumn,showpacs,superscriptaddress,longbibliography,preprintnumbers,floatfix,amssymb,amsmath]{revtex4-2}
\makeatletter
\newcommand\showtitleinbib{{\escapechar=`\\ \immediate\write\@auxout{%
\csname citation{REVTEX42Control}\endcsname^^J%
\csname citation{apsrev42Control}\endcsname
}}}
\makeatother
\usepackage{graphicx}  
\usepackage{dcolumn}   
\usepackage{bm}        

\usepackage{caption}
\DeclareCaptionJustification{justified}{\leftskip=0pt\rightskip=0pt\parfillskip=0ptplus1fil\relax}
\usepackage{subcaption}
\usepackage{setspace}

\usepackage{color}
\usepackage[normalem]{ulem}
\newif\ifshowedits
\showeditstrue
\newcommand{\tanmoy}[2][]{{\ifshowedits\color{red}\ifx\undefined#1\undefined\else\sout{#1}\fi\fi
#2}}
\newcommand{\junsik}[2][]{{\ifshowedits\color{red}\ifx\undefined#1\undefined\else\sout{#1}\fi\fi
#2}}
\newcommand{\FIXME}[1]{[\textcolor{red}{FIXME: #1}]}

\begin{document}
\title{Electroweak box diagram contribution for pion and kaon decay from lattice QCD}

\author{Jun-Sik Yoo}%
\email{junsik@lanl.gov}

\author{Tanmoy Bhattacharya}
\email{tanmoy@lanl.gov}
\author{Rajan Gupta}%
\email{rajan@lanl.gov}
\affiliation{Los Alamos National Laboratory, Theoretical Division T-2, Los Alamos, New Mexico 87545, USA}
\author{Santanu Mondal}%
\email{santanu.sinp@gmail.com}
\affiliation{Department of Physics and Astronomy, Michigan State University, MI, 48824, USA}
\affiliation{ArcVision Technologies, 34 Mahatma Gandhi Road, Kolkata 700009, India}
\author{Boram Yoon}%
\email{byoon@nvidia.com}
\affiliation{NVIDIA Corporation, Santa Clara, CA 95051, USA}

\date{\today}
\preprint{LA-UR-23-24250}

\begin{abstract}
One of the sensitive probes of physics beyond the standard model is the test of the unitarity of the Cabbibo-Kobyashi-Maskawa (CKM) matrix. Current analysis of the first row is based on $|V_{ud}|$ from fourteen superallowed $0^+\!\! \to  0^+ $ nuclear $\beta$ decays and $|V_{ud}|$ from the kaon semileptonic decay, $K \to \pi \ell \nu_\ell$. Modeling the nuclear effects in the $0^+\!\! \to  0^+ $ decays is a major source of uncertainty, which would be absent  in neutron decays. To make neutron decay competitive requires improving the measurement of neutron lifetime and the axial charge, as well as the calculation of the radiative corrections (RC) to the decay. The largest uncertainty in these RCs, which comes from the non-perturbative part of the $\gamma W$-box diagram and its evaluation using lattice QCD, is still not under control. Here, we show that the analogous calculations for the pion and kaon decays are robust and give $\square_{\gamma W}^{VA} |_{\pi} = 2.810 (26) \times 10^{-3} $ and 
$ \square_{\gamma W}^{VA} \Big|_{K^{0, S U(3)}} = 2.389 (17) \times 10^{-3}$ in agreement with the previous analysis carried out by Feng et al. using a different discretization of the fermion action. 
 \end{abstract}

\maketitle

\section{Introduction}

\begin{figure}  
    \centering
    \includegraphics[width=0.52\textwidth]{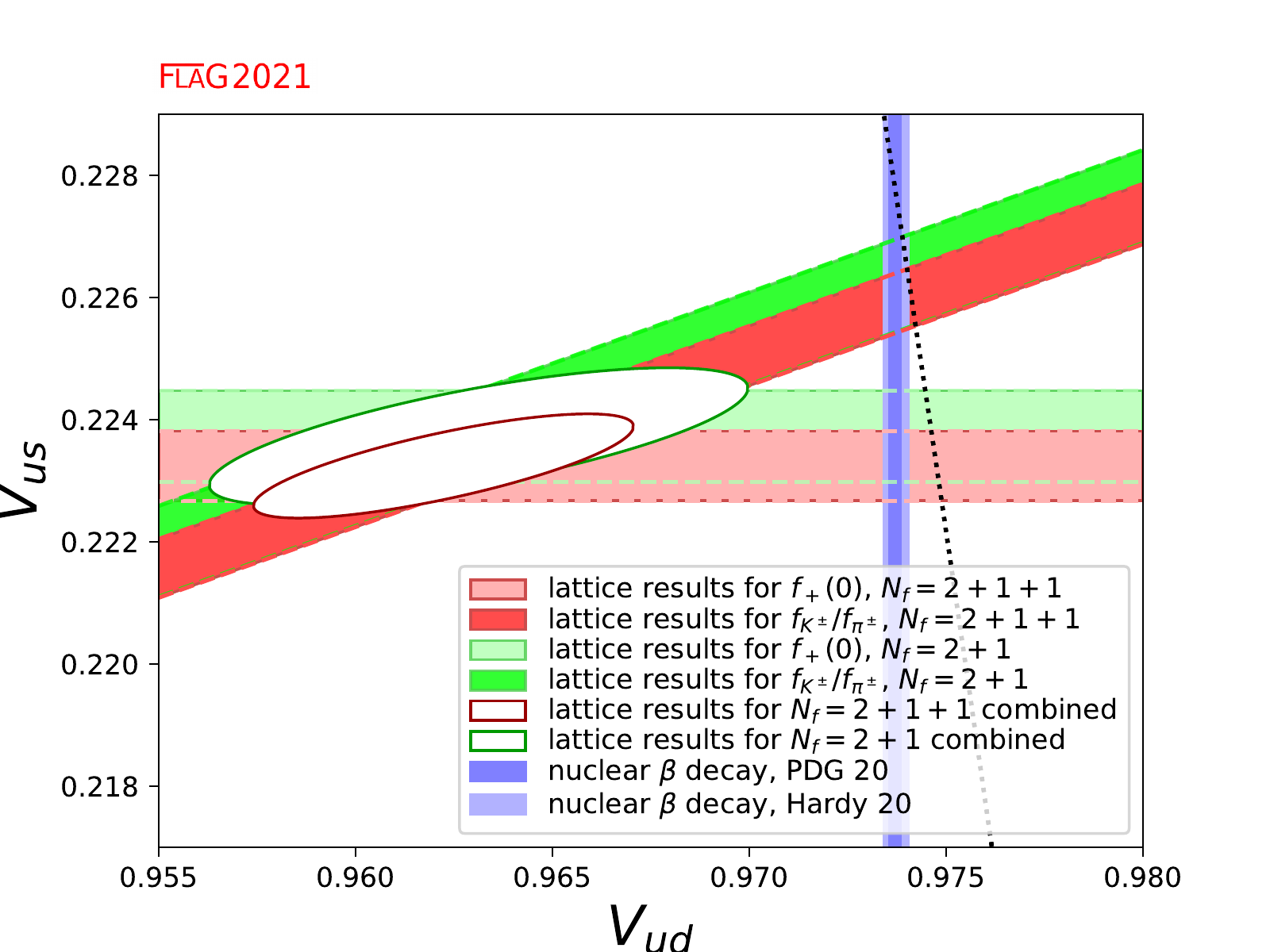}   
    \caption{Current status of $|V_{ud}|$ and $|V_{us}|$ using 2+1+1 and 2+1 flavor lattice calculations combined with nuclear $\beta$ transitions (vertical blue bands). The black dotted line gives the correlation between $|V_{ud}|$ and $|V_{us}|$ if the CKM matrix is unitary. Figure reproduced from FLAG report 2021~\cite{Aoki:2021kgd}}
    \label{fig:FLAG21}
\end{figure}
 
\begin{figure}[ht] 
    \centering
    \includegraphics[width=.48\textwidth]{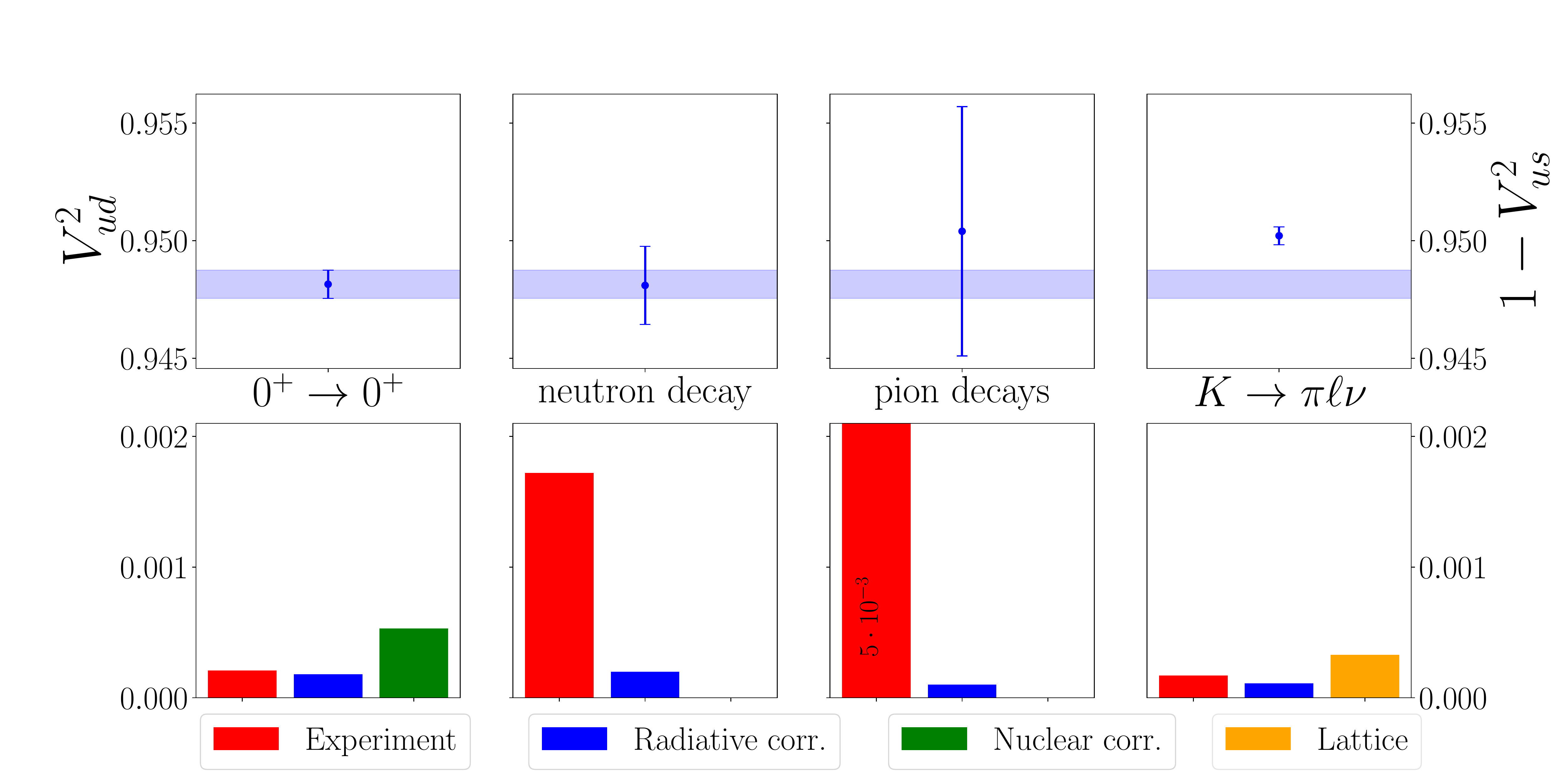}
    \caption{The values (top) and the error budgets (bottom) of three extractions of $|V_{ud}|^2$ and one of $|V_{us}|^2$ to test the unitarity of the first row of the CKM matrix \cite{Workman:2022ynf,UCNt:2021pcg,Hardy_2012}. }
    \label{fig:Errorbudget}
\end{figure}

In the intensity frontier, physics beyond the standard model (BSM) is probed by confronting accurate predictions of the standard model (SM) with precision experiments. Today, there are several tests showing roughly 2--3$\sigma$ deviations, one being the unitarity of the first row of the CKM quark mixing matrix, which states that $\Delta_{\textrm{CKM}} \equiv\allowbreak |V_{ud}|^2+\allowbreak|V_{us}|^2+\allowbreak|V_{ub}|^2 - 1$ should be zero. There is a {$\approx 3\sigma$} tension with the SM~\cite{Workman:2022ynf,Seng:2018yzq,Seng:2018qru,Czarnecki:2019mwq} in 
current analyses using the most precise value of $|V_{ud}|^2=0.94815(60)$ coming from $0^+\!\! \to  0^+ $   nuclear $\beta$ decays \cite{Workman:2022ynf}, and  $|V_{us}|^2=0.04976(25)$ obtained from kaon semileptonic decays ($K \to \pi \ell \nu_\ell$) along with the $N_f = 2+1+1$-flavor lattice result for $f^K_+(0)$~\cite{Aoki:2021kgd}. The estimate of  $|V_{ub}|^2\approx(2\pm 0.4) \times 10^{-5}$ is too small to impact the unitarity test. 

A current analysis of the unitarity bound is shown in Fig.~\ref{fig:FLAG21}, with the errors from various sources in $0^+\!\! \to  0^+ $ nuclear, nucleon, pion and kaon decays shown in Fig.~\ref{fig:Errorbudget}. While the extraction of $V_{ud}$ from superallowed $0^+ \rightarrow 0^+$ nuclear decays is the best, it is still subject to significant uncertainty in the theoretical analysis of nuclear effects.  

\begin{figure}[ht]  
\includegraphics[viewport=0 85 792 527,clip,width=0.52\textwidth]{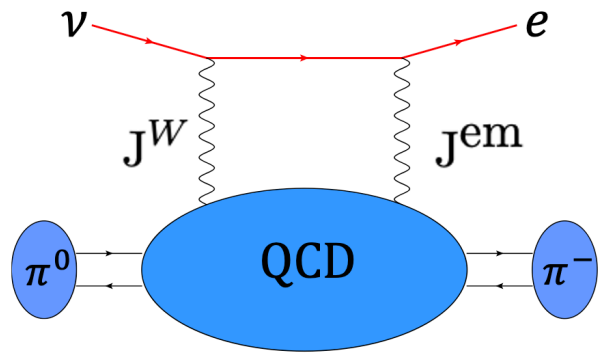}
\caption{Schematic of the $\gamma W$-box diagram that needs to be calculated to obtain the RC to pion decay.} 
\label{fig:box_diagram_pion}
\end{figure}

Theoretically, the neutron is a clean system, i.e., 
there are no nuclear corrections to consider. The largest theoretical uncertainty, as discussed in Refs.~\cite{Sirlin:1977sv,Feng:2020zdc,Ma:2021azh}, comes from the radiative corrections (RC) given by the $\gamma W$-box diagram illustrated in Fig.~\ref{fig:box_diagram_pion} for the pion. Lattice QCD is the best method to determine the non-perturbative part of the $\gamma W$-box for all three (neutron, pion and kaon) $\beta$ decays of interest. For the neutron, this, together with improvements in experiments measuring free neutron lifetime, $\tau_n$, and the axial charge, $g_A$, will make the extraction of $|V_{ud}|^2$ competitive with that from  $0^+ \rightarrow 0^+$ nuclear decays and have the advantage of no nuclear corrections.\looseness-1

In this paper we present results for the simpler cases of the \(\beta\) decay of the pion and the kaon as we have not yet obtained a signal in the neutron correlation functions. Nevertheless, we provide a brief review of the status of the extraction of $|V_{ud}|^2$ from neutron decay as it is the ultimate goal of this project.  The analysis is carried out using the
 formula~\cite{Czarnecki:2019mwq,Czarnecki:2018okw}
\vspace{-6pt}\begin{align}
\left|V_{ud}\right|^2 &= \left(  \frac{G_\mu^2 m_e^5 }{2\pi^3} f \right)^{-1} \frac{1}{ \tau_n (1 + 3 g_A^2) (1 + \textrm{RC}) }
\nonumber\\&= \frac{5099.3(3) \ \textrm{seconds}}{\tau_n (1 + 3 g_A^2) (1 + \textrm{RC})}\label{master}
\end{align}
where  $g_A$ is best 
obtained from the neutron $\beta$ decay asymmetry parameter $A$,
$G_\mu$ is the Fermi constant extracted from muon decays, and
$f=1.6887(1)$ is a phase space factor. With future measurements of the neutron lifetime $\tau_n$ reaching an uncertainty of $\Delta\tau_n \sim 0.1$s, and of the ratio $\lambda = g_A/g_V$ of the neutron axial and vector coupling reaching  $\Delta\lambda/|\lambda|\sim 0.01\%$, the extraction of $V_{ud}$ with accuracy comparable to $0^+\!\! \to  0^+ $ superallowed $\beta$ decay can be achieved provided the uncertainty in the RC to neutron decay can be reduced. 

The lattice methodology for the calculation of RC to pion, kaon and neutron decays (the $\gamma W$-box diagram illustrated in Fig.~\ref{fig:box_diagram_pion} for the pion) is similar~\cite{Feng:2020zdc,Ma:2021azh}. From here on we restrict the discussion to  pion and kaon semileptonic decays, for which  the analogues of Eq.~\eqref{master} to extract $|V_{ud}|^2$ and $|V_{us}|^2$ are~\cite{Cirigliano:2002ng,Pocanic:2003pf,Czarnecki:2019iwz,Workman:2022ynf}
\begin{align}
    |V_{ud}f_+^\pi(0)|^2_{\pi\ell}\!=& \frac{64\pi^3  \,\Gamma_\pi}{G_\mu^2 M_\pi^5 I_{\pi} \left( 1 + \delta  \right) }\\
    |V_{us}f_+^K(0)|^2_{K\ell}\!=& \frac{192\pi^3 \,\textrm{BR}(K\ell ) \,\Gamma_K}{G_\mu^2 M_K^5 C_K^2 S_{EW} I_{K\ell} \left( 1 + \delta^{K\ell}_{EM} + \delta^{K\ell}_{SU(2)} \right) } \,,
\end{align}
where $\Gamma_{\pi/K}$ are the $\pi$ and K decay rates, $I_{\pi,K}$ are known kinematic factors, $f_+^{\pi/K}$ are semileptonic form factors, $C_K$ is a known normalization factor needed for kaon decay, $S_{EW}$ is the short distance radiative correction, and the $\delta^{K\ell}_{SU(2)} $ is the isospin breaking correction. The two (long distance) radiative corrections in which the uncertainty needs to be reduced are $\delta$ for pion and  $\delta^{K\ell}_{EM}$ for kaon decay.  

Looking ahead, the experimental uncertainty in pion decay needs  to be reduced by a factor greater than 20, at which point it will become  roughly equal to that in radiative corrections. PIONEER~\cite{PIONEER:2022yag} is a next generation experiment aimed at measuring the rare pion decay branching ratios precisely. Its primary goal is to improve the measurement of the branching ratio of the semileptonic decay  by up to a factor of ten, thus  reducing the experimental uncertainty in $|V_{ud}|^2$ by the same factor. At that point, as shown in Fig.~\ref{fig:Errorbudget}, the experimental error in $|V_{ud}|^2$ from pion decay will become comparable  to that from $0^+ \!\!\rightarrow 0^+$ superallowed nuclear decay, and have a small theory uncertainty.

In the determination of $|V_{us}|$ from kaon $\beta$ decay, 
the largest uncertainty comes from $f_+(0) $ taken from lattice calculations~\cite{FlavourLatticeAveragingGroupFLAG:2021npn}.  Comparatively, the uncertainty in the radiative correction and experiments is already small~\cite{Seng:2021nar}.

This paper is organized as follows. The essential formulae needed to describe the calculation are summarized in the next section~\ref{sec:EWbox}. 
The lattice setup is given in Sec.~\ref{sec:LQCD}, error reduction methods used in the extraction of the hadronic tensor $\mathcal{H}_{\mu \nu}^{V A}$ in Sec.~\ref{sec:err_red}, a comparison of results for $\mathcal{M}_H(Q^2)$  with perturbation theory in Sec.~\ref{sec:pert}, and the extrapolation to the continuum limit in Sec.~\ref{sec:CCFV}.  The final results and the comparison to previous calculations are given in Sec.~\ref{sec:results}. 

\section{Electroweak Box Diagram}
\label{sec:EWbox}

\begin{figure}[ht]  
    (A)\includegraphics[width=0.40\linewidth]{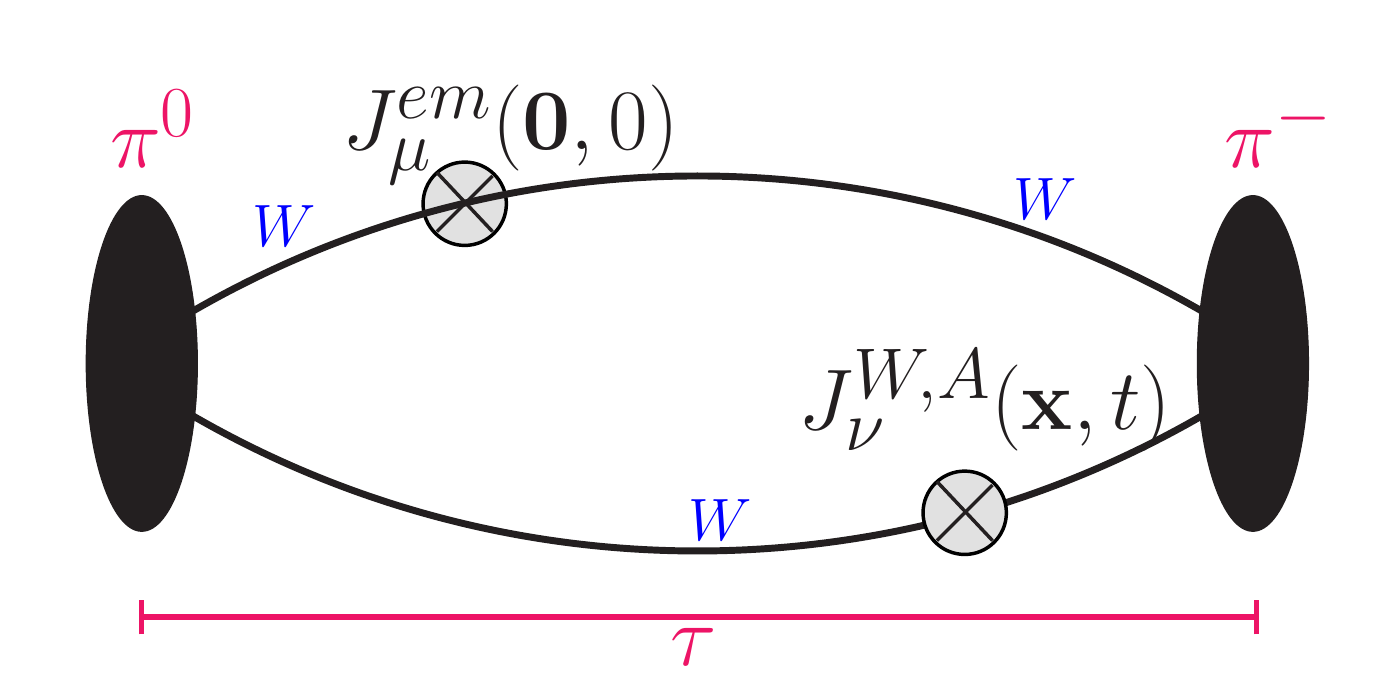} 
    (B)\includegraphics[width=0.44\linewidth]{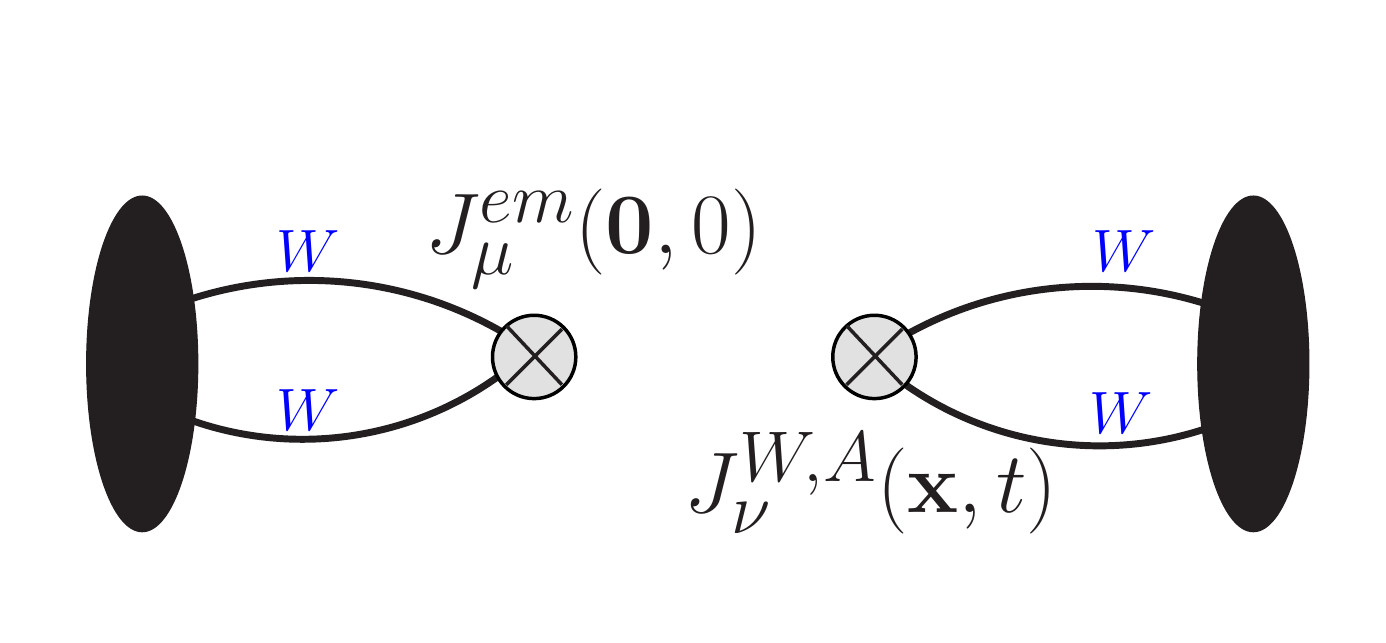} \\
    (C)\includegraphics[width=0.42\linewidth]{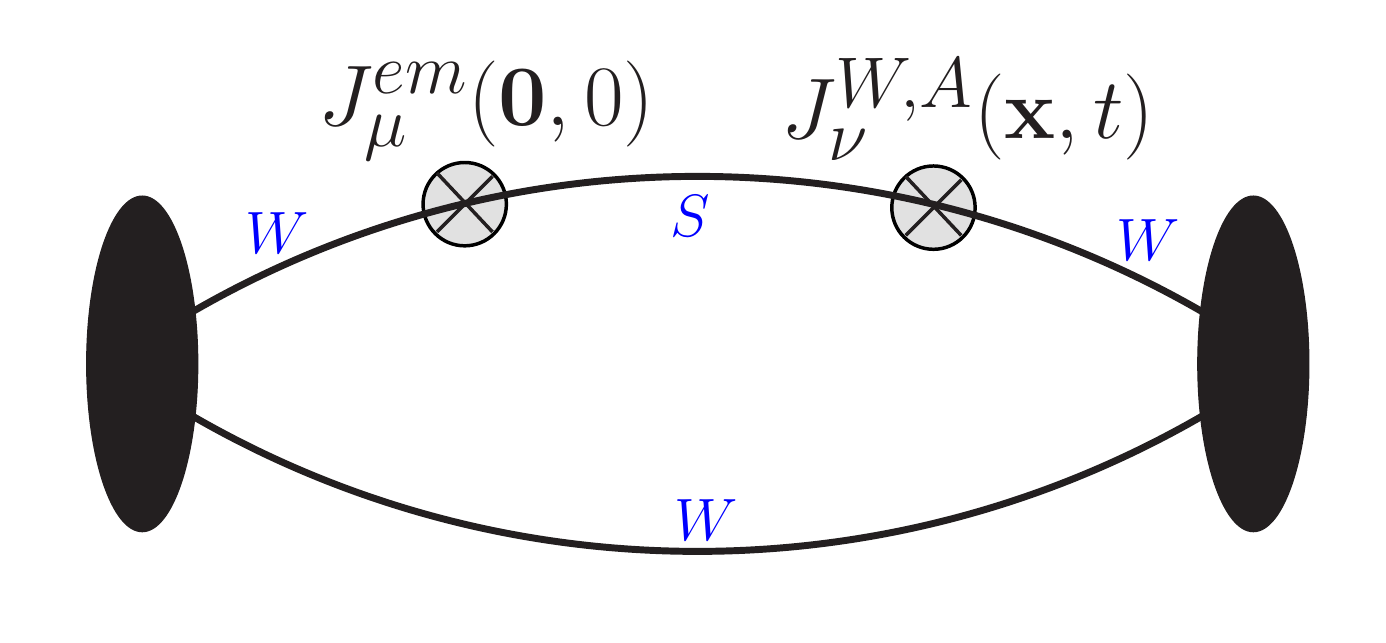} 
    (D)\includegraphics[width=0.42\linewidth]{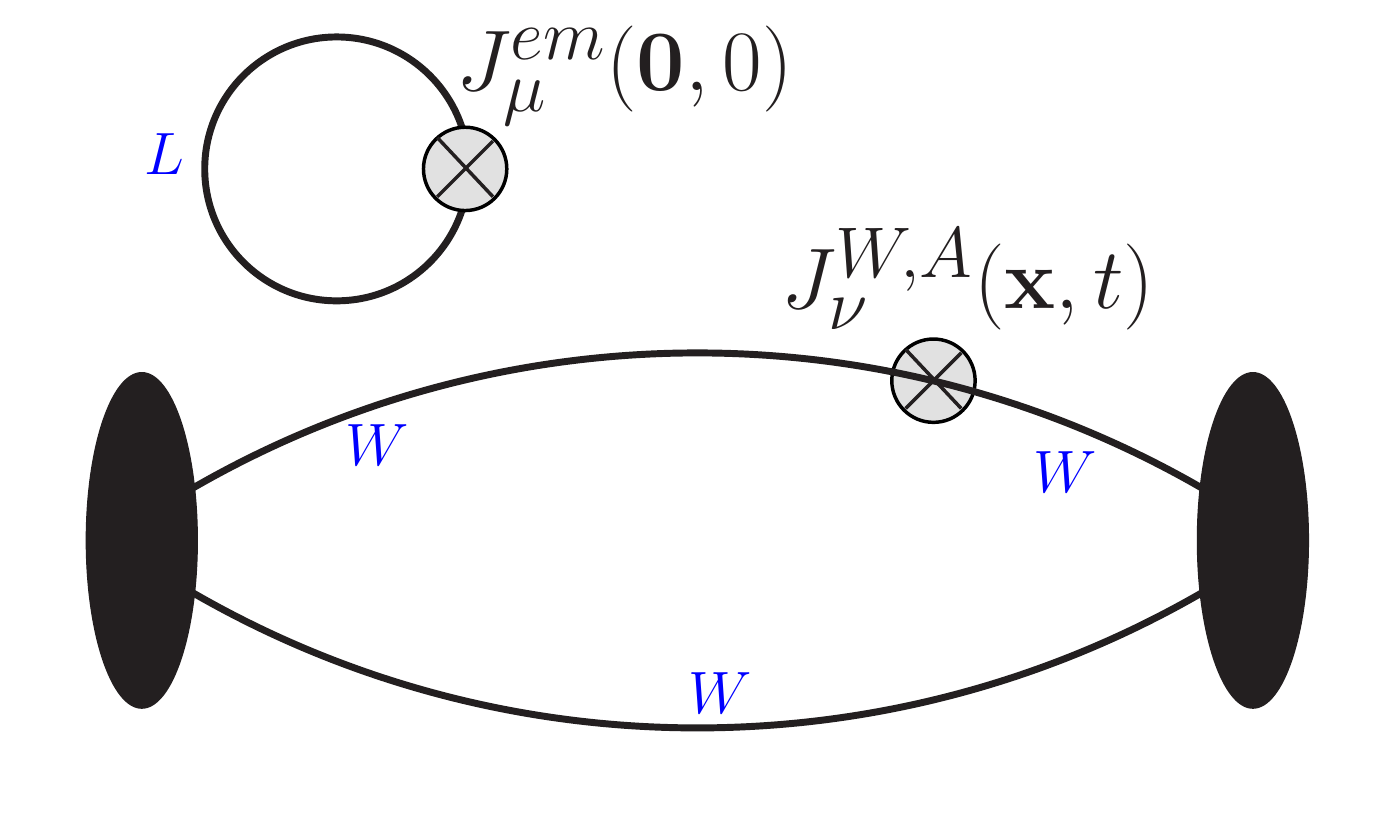}
\caption{The four quark-line diagrams that contribute, in general, to $\mathcal{H}_{\mu \nu}^{V A}(\vec{x},t) = \langle \pi | \textrm{T} [J_\mu^{em}(x) J_\nu^{W,A}(0)] | \pi \rangle $ in the pion $\gamma W$-box.}
\label{fig:diagrams}
\end{figure}

Following the framework developed in \cite{PhysRevD.100.013001,Feng:2020zdc}, the calculation of the electroweak box diagram requires evaluating the four quark line diagrams shown in Fig.~\ref{fig:diagrams}. The result can be written as
\begin{align}\label{delta}
    \left.\square_{\gamma W}^{V A}\right|_H  &= \int_0^{+\infty} dQ^2
    \int^{ Q}_{-  Q}  
     d Q_0 \frac{1}{Q^4} \frac{1}{Q^2 + M_W^2} \times{}\nonumber\\
     &\qquad\qquad\qquad L^{\mu\nu}(Q,Q_0) T^{VA}_{\mu\nu}(Q,Q_0)\ 
\end{align}
with $H$ labeling the hadron ($\pi,\ K$, or $ N$) with 
mass $M_H$ under consideration, and $M_W$ the $W$ meson mass.
Substituting in the known leptonic part $L^{\mu\nu}(Q,Q_0)$ gives 
\begin{equation}
\begin{aligned}
\left.\square_{\gamma W}^{V A}\right|_H= & -\frac{1}{F_{+}^H} \frac{\alpha_e}{\pi} \int_0^{\infty} d Q^2 \frac{M_W^2}{M_W^2+Q^2} \\
& \times \int_{-\sqrt{Q^2}}^{\sqrt{Q^2}} \frac{d Q_0}{\pi} \frac{\left(Q^2-Q_0^2\right)^{\frac{3}{2}}}{\left(Q^2\right)^2}\frac{\epsilon_{\mu \nu \alpha \beta} Q_\alpha P_\beta T_{\mu \nu}^{V A}}{2 M_H^2|\vec{Q}|^2} .
\end{aligned}
\end{equation}
The hadronic tensor $T^{VA}_{\mu\nu}$ is given by 
\begin{align}
T^{VA}_{\mu\nu}&=\frac{1}{2}\int d^4x\,e^{iQ \cdot x}\  \mathcal{H}_{\mu \nu}^{V A}(\vec{x},t) \,,
\label{eq:Tmunu}
\end{align}  
with 
\begin{equation}
\mathcal{H}_{\mu \nu}^{V A}(\vec{x},t) \equiv  \langle
H_f (p)|T\left[J^{em}_\mu(0,0)\,J^{W,A}_\nu({\vec x},t)\right]|H_i(p)\rangle \,, 
\label{eq:Hmunu}
\vspace{4pt}
\end{equation}
where   $J^{W,A}_\mu = Z_A \bar{u} \gamma_\mu \gamma_5 d $ and $J_\mu^{em} = Z_V ( \frac{2}{3}\bar{u}  \gamma_\mu d - \frac{1}{3}\bar{d}  \gamma_\mu d)$ are the renormalized  axial (A) and vector (V) currents with $Z_A$ and $Z_V$ calculated in Ref.~\cite{Gupta:2018qil}. The Wick contractions in the hadronic part, $\mathcal{H}_{\mu \nu}^{V A}(\vec{x},t)$, give rise to, in general, the four types of  quark-line diagrams shown in Fig.~\ref{fig:diagrams} for pion decay, and is the quantity we calculate on the lattice. 

Only one term, $T_3$, in the spin-independent part of the expansion $T^{VA}_{\mu\nu}= i \epsilon_{\mu \nu \alpha \beta} q^\alpha p^\beta T_3 +
\dots$  contributes~\cite{Seng:2018qru,Feng:2020zdc}. Knowing $T_3$ as a function of $Q^2$, the $\gamma W$-box correction, using the notation in Refs.~\cite{Feng:2020zdc,Ma:2021azh}, is given by
\begin{equation}
\Box^{VA}_{\gamma W} = \frac{3 \alpha_e}{2 \pi} \int \frac{dQ^2}{Q^2}
\frac{M_W^2}{M_W^2+Q^2} {\mathcal{ M}_H(Q^2)}
\label{eq:boxMH}
\end{equation}
with
\begin{align}
    {\mathcal{M}_{H}\left(Q^{2}\right)}&=-\frac{1}{6} \frac{1}{F_{+}^{H}} \frac{\sqrt{Q^{2}}}{M_{H}} \int d^{4} x\, \omega(\vec{x},t) \times{}\nonumber\\
    &\qquad\qquad\qquad\epsilon_{\mu \nu \alpha 0} \,x_{\alpha} \,\mathcal{H}_{\mu \nu}^{V A}(\vec{x},t) \,, \label{eq:eq9}\\
    \omega(t, \vec{x}) = &\int_{-\frac{\pi}{2}}^{\frac{\pi}{2}} \frac{\cos ^3 \theta d \theta}{\pi} \frac{j_1\left(\sqrt{Q^2}|\vec{x}| \cos \theta\right)}{|\vec{x}|} \times{}\nonumber\\
    &\qquad\qquad\qquad\cos \left(\sqrt{Q^2} t \sin \theta\right) \,,  
\label{eq:MHdef}
\end{align}
and where $j_1$ in the weight function $\omega(t, \vec{x})$ is the spherical Bessel function. While $\mathcal{H}_{\mu \nu}^{V A}(\vec{x},t)$ is a function of the separation $\{\vec x, t\}$ and, on the lattice, the integral in Eq.~\eqref{eq:eq9} becomes a sum over the discrete values of these coordinates, ${\mathcal{M}_{H}\left(Q^{2}\right)}$ is, however, available for all values of $Q^2$ as can be seen from Eq.~\eqref{eq:MHdef}. 

\begin{figure}[ht]   
    \centering
    \includegraphics[width=.50\textwidth]{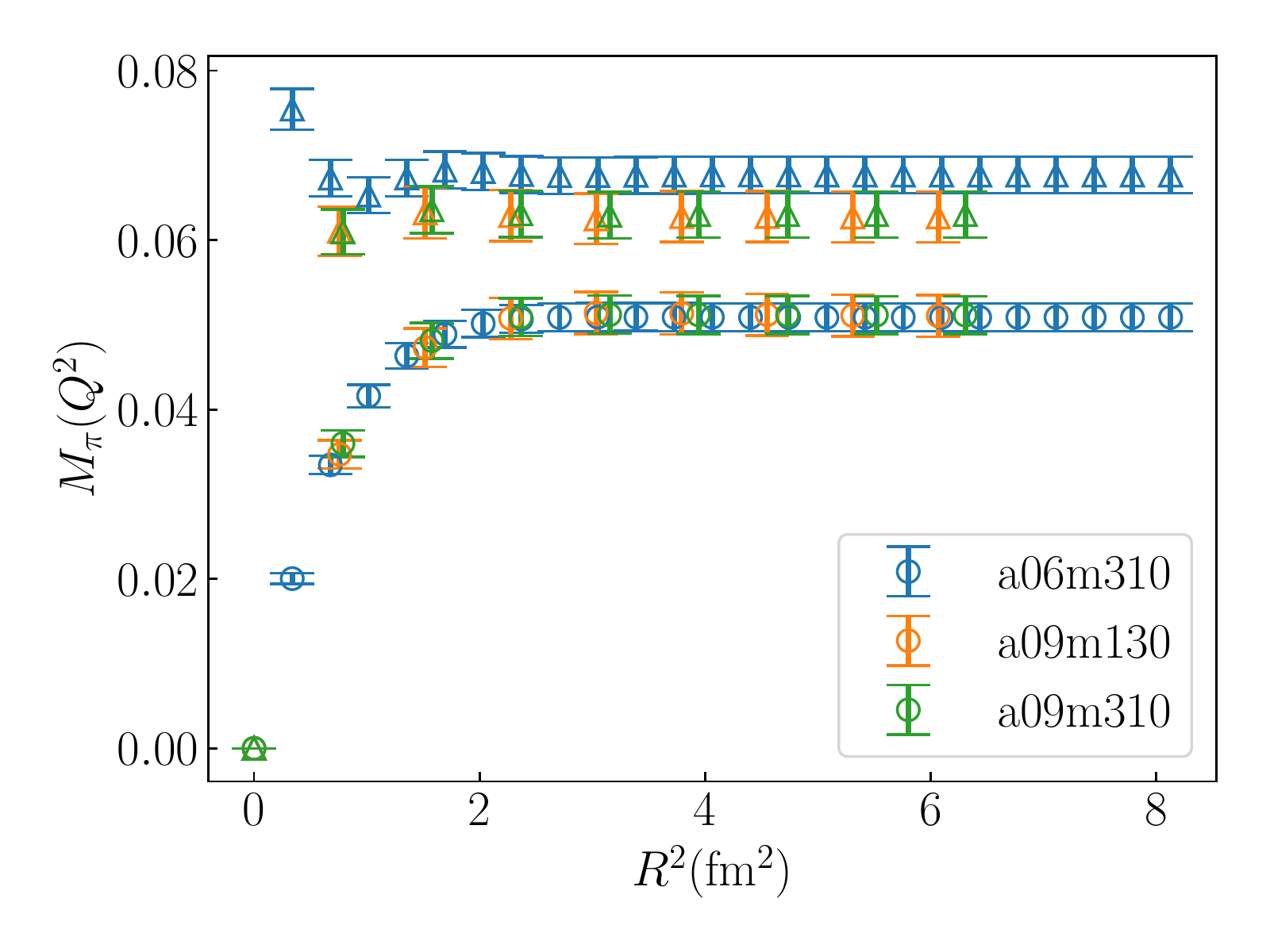}
    \vspace{-0.2cm}
    \caption{Check of the convergence of ${\cal M}_\pi$ versus $R^2 \equiv |\vec{x}|^2 + t^2$ as defined in Eq.~\protect\eqref{eq:eq9}. Circles are used for $Q^2 = 0.317$ GeV$^2$ and triangles for $Q^2 = 3.0$ GeV$^2$ data.\looseness-1}
    \label{fig:Rsq-and-ratio}
\end{figure}

One expects the signal in $\mathcal{H}_{\mu \nu}^{V A}(\vec{x},t)$ to fall off with $\{\vec x, t\}$, and in Fig.~\ref{fig:Rsq-and-ratio}, we show that the integral saturates for $ R^2 \gtrsim 2\ \textrm{fm}^2$. To be conservative and yet save computation time,  we choose the integration volume to be smaller than the entire lattice but larger than $ R^2 \sim 3.3\ \textrm{fm}^2$ on all the ensembles. 

\vspace{.5cm}

\section{Lattice Setup}
\label{sec:LQCD}

\begin{figure}[ht]  
    \includegraphics[width=.50\textwidth]{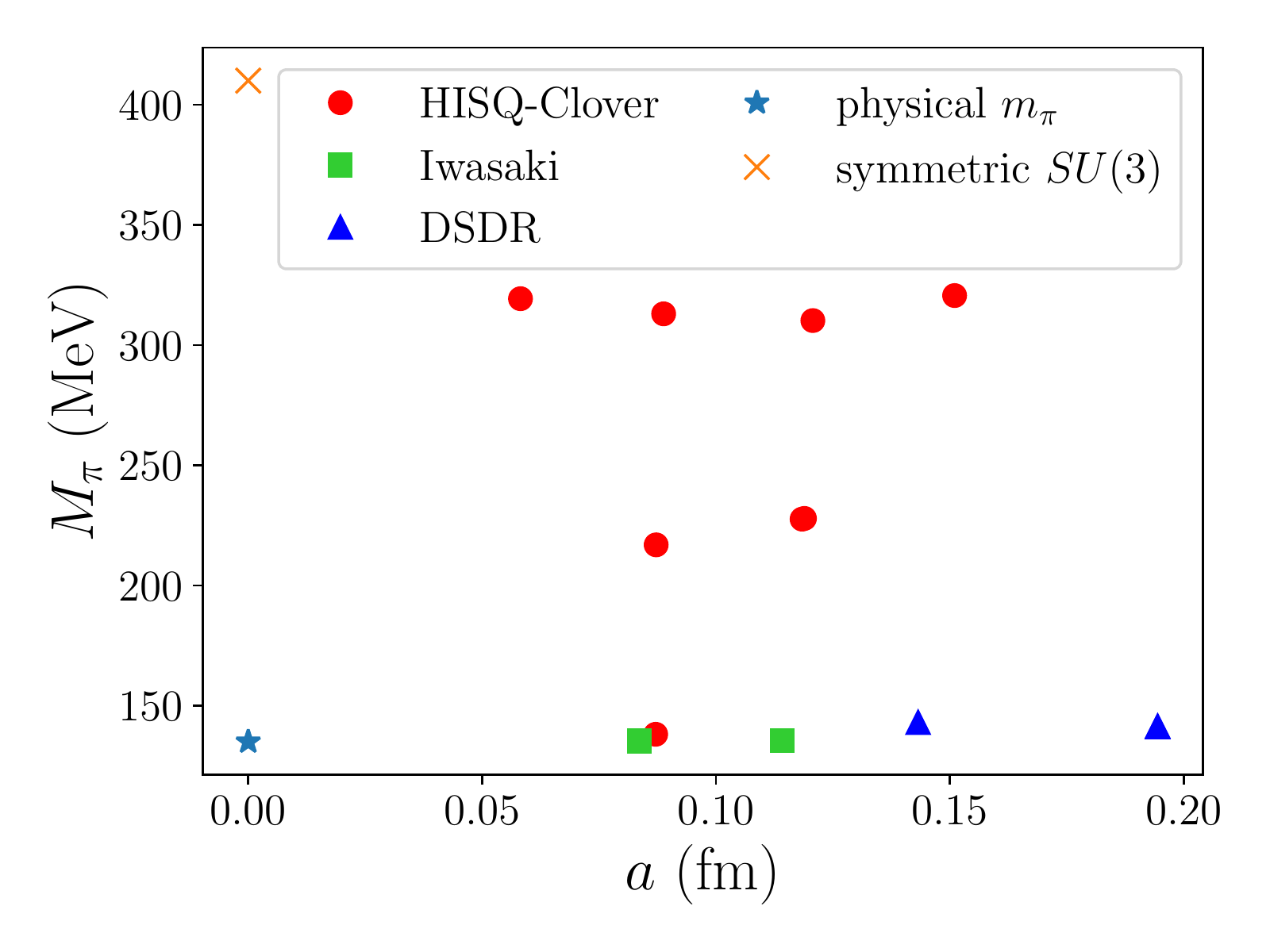}
    \caption{The lattice spacing and pion mass of the eight  ensembles (red circles) with 2+1-flavors of Wilson-clover fermions analyzed in this study. The physical point (Star) and SU(3) symmetric point (Cross) are also shown. We also show the points for  DSDR (blue triangle) and Iwasaki (green square) actions that were used in Feng~et al. \cite{Feng:2020zdc,Ma:2021azh}. }
    \label{fig:latt_param}
\end{figure}

We have performed the calculation using eight $N_f = 2+1+1$-flavor HISQ sea quark ensembles generated by the MILC collaboration~\cite{Bazavov:2012xda}, whose parameters are given in Table~\ref{tab:ensembles}, and shown in the $\{a,M_\pi\}$ plane in Fig.~\ref{fig:latt_param}. For comparison, we also show the parameters in the ``Iwasaki'' and ``DSDR'' variants of domain-wall fermions used in Ref.~\cite{Feng:2020zdc}.

The correlation functions corresponding to the quark-line diagrams in Fig.~\ref{fig:diagrams} are constructed using Wilson-clover fermions, and  the tuning of the light quark mass in the isosymmetric limit  is done by requiring $M_\pi^{\rm valence} = M_\pi^{\rm sea}$ as described in Ref.~\cite{Gupta:2018qil}. The strong coupling, $\alpha_s(1/a)$ in $\overline{\rm MS}$ scheme, for each lattice ensemble was computed using the fourth order perturbative expression \cite{Deur:2016tte} with $\Lambda_{\text{QCD}}^{n_f = 4} = 292\ \textrm{MeV}$ taken from \cite{ParticleDataGroup:2016lqr}. 

\begin{figure}[ht] 
    \includegraphics[width=.50\textwidth]{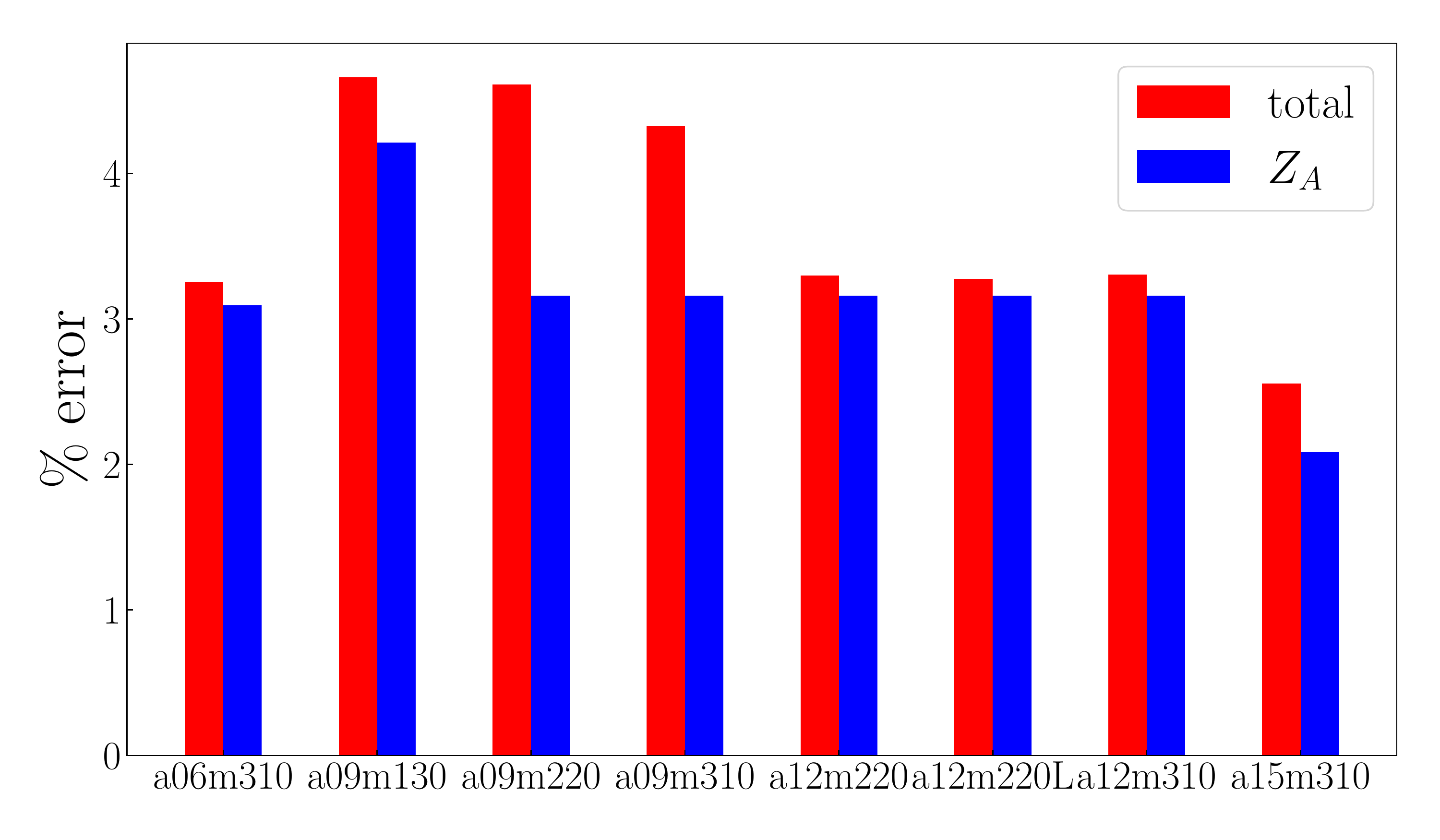}
    \caption{Fractional error in the calculation of the box diagram for the pion on the eight ensembles. The total  uncertainty (red bar) in the calculation is dominated by the uncertainty from the renormalization constant $Z_A$ (blue bar).}
    \label{fig:error_budget}
\end{figure}

Of the four quark-line diagrams shown in Fig.~\ref{fig:diagrams}, diagrams are A and C are called ``connected''. The ``disconnected'' diagram (B) does not contribute due to the $\gamma_5-$hermiticity property of the quark propagator, and diagram (D) vanishes in the SU(3) limit and is not evaluated here. To construct these correlation functions, quark propagators are generated with wall sources at two ends of a sublattice with separation $\tau$ in time (see Table~\ref{tab:ensembles}). We label these quark lines by W. For the internal line S in diagram C, we solve for an additional propagator from the position of the vector current $V_\mu$ inserted on the middle timeslice between the source and sink. This point is labeled $\{{\vec x}=0, t=0\}$. We choose 256 such points for diagram A and 64 for diagram C. Data are collected with the position of $A_\mu$ varied within distance $R^2$, listed in Table~\ref{tab:ensembles}, from these points. 
On each configuration, we use 8 regions (sublattices) offset by $N_T / 8$ on which we repeat the calculation to further increase the statistics. (On the physical pion mass ensemble a09m130, we double the number to 16 regions.)  With the current statistics, the errors in the data from the eight ensembles are comparable as shown later in Fig.~\ref{fig:mass_dep}. Since the total error budget for the box diagram is already dominated by the uncertainty in the renormalization constant $Z_A$ as shown in Fig.~\ref{fig:error_budget}, the 
current statistics are considered sufficient.

\begin{figure}[ht]  
      \includegraphics[width=.49\textwidth]{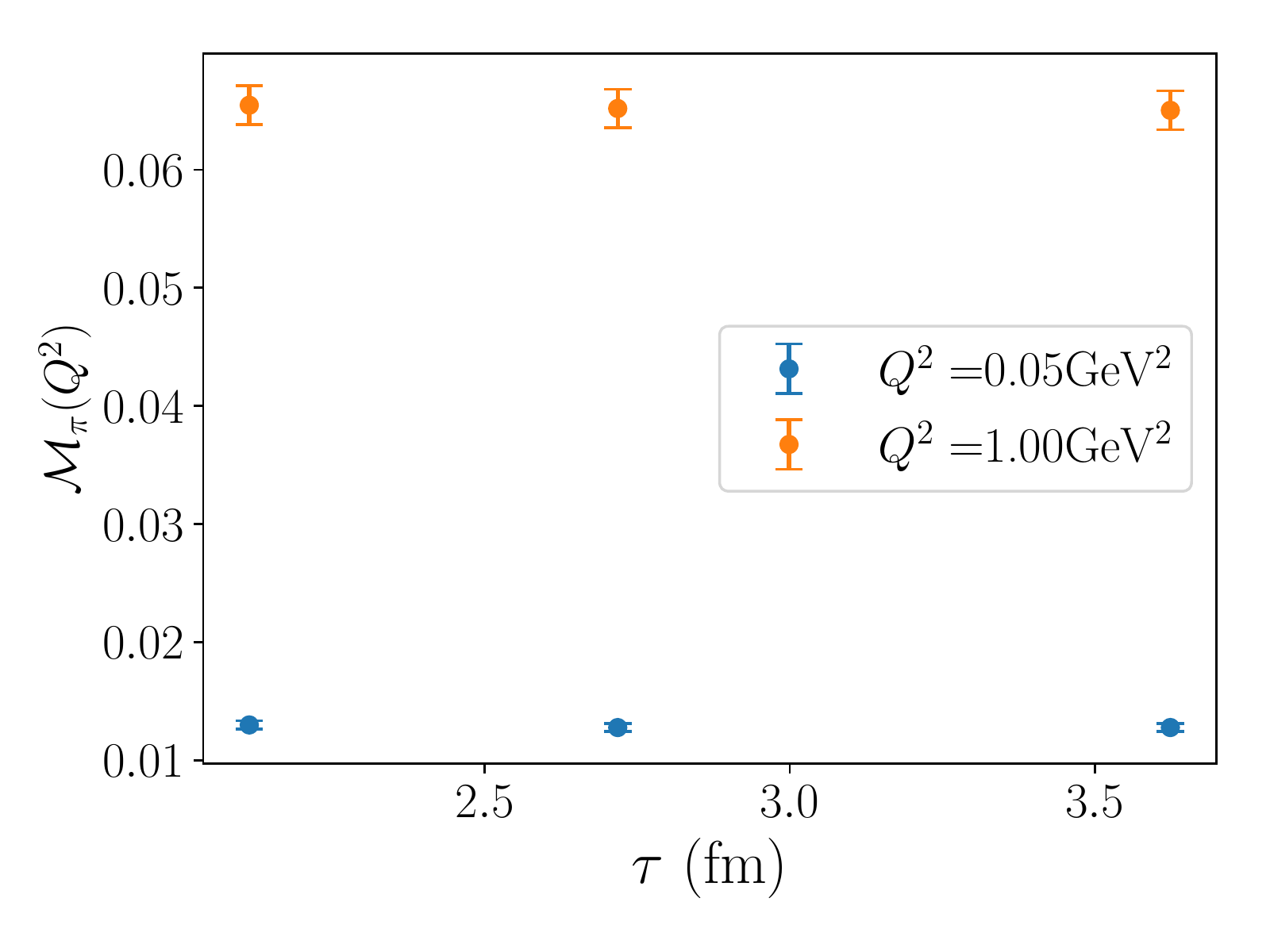}\\
    \includegraphics[width=.49\textwidth]{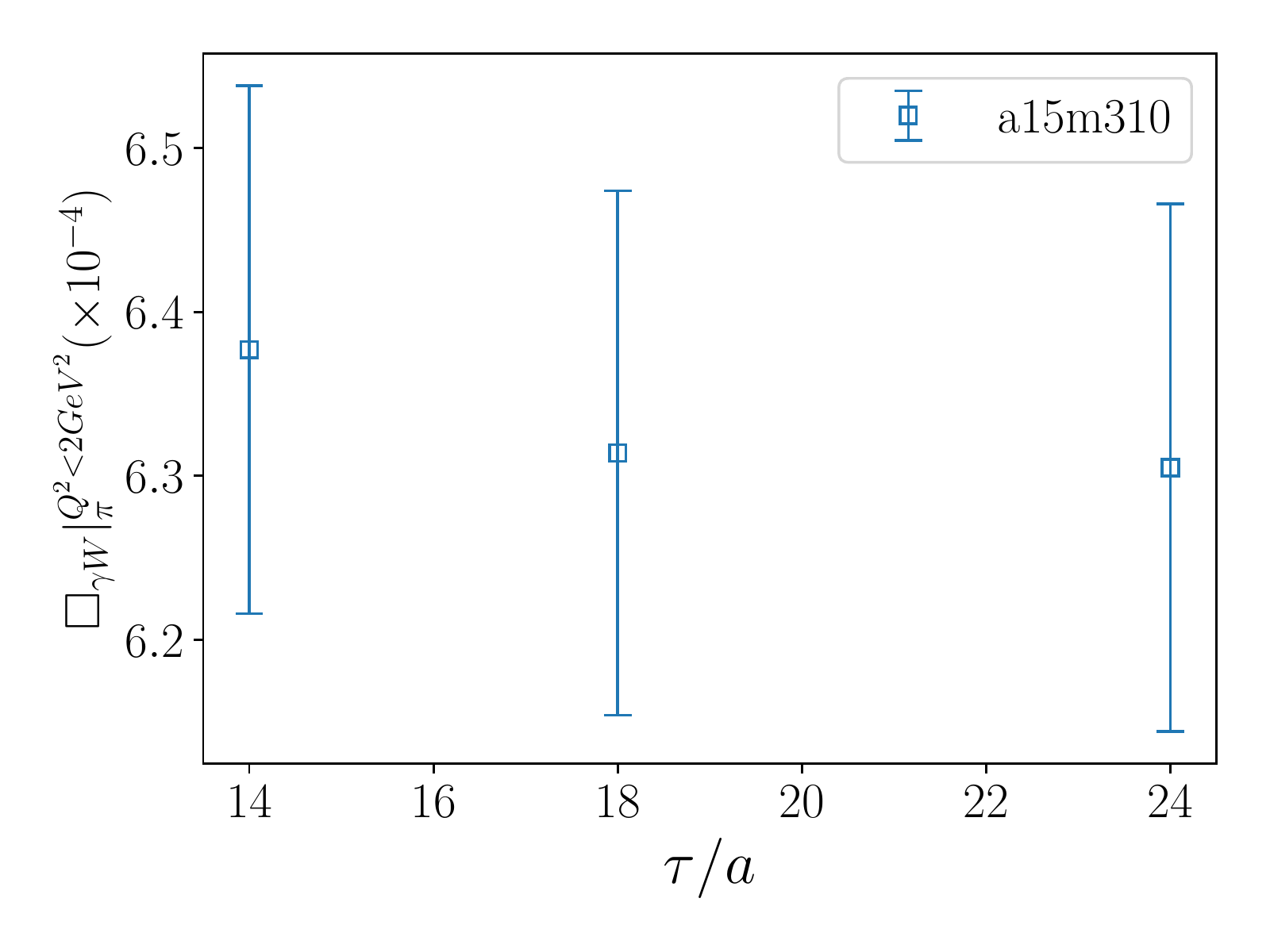}
    \caption{The data for $\mathcal{M}_{\pi}(Q^2)$ and the $\gamma W$-box contribution on the $a15m310$ ensemble show no significant dependence on the source-sink separation $\tau$.  We chose to perform all analyses in this paper with $\tau$ in the range $[3.48 \le \tau \le 3.63] $~fm.  }
    \label{fig:tsep_dep}
\end{figure}

In Fig.~\ref{fig:tsep_dep}, we show the result for ${\cal M}_\pi$ and the $\gamma W$-box as a function of 
the separation $\tau$ between the wall source and the sink.  Our data show no significant dependence on $\tau$ for $\tau > 2.4$~fm, at which separation the contribution is deemed dominated by the ground state. To be conservative and  because the signal in correlation functions for pseudoscalar mesons does not degrade with $\tau$, we chose to work with larger values of $\tau$ in the range $[3.48 \le \tau \le 3.63] $~fm on all ensembles. Note that this ability to choose $\tau$ large enough to isolate the ground state is special to pseudoscalar mesons. For our target case of neutrons, the signal decays exponentially and excited state contamination is a severe challenge~\cite{Park:2021ypf}. As a result, even with much larger statistics, our ongoing calculations for neutrons have not yielded a statistically significant signal.

\begin{table*}[t]
\tabcolsep 6pt
    \begin{center}
    \begin{tabular}{c|c|c|c|c|c|c|c|c|c|c}
    \hline \hline 
        Ensemble ID & a & $\alpha_S$ & $M_\pi^{val}$ & $M_\pi^{sea}$ & $L^3 \times T$ & $m_\pi L$ & $\tau / a$ & $(R/a)^2$ & $R^2 $ & $N_{conf}$  \\
                 & [fm] &  & [MeV] & [MeV]&  &  &  &  &  $[\text{fm}^2]$ &   \\
        \hline 
        a06m310 & 0.0582(04) & 0.2433 &319.6(2.2) & 319.3(5) & $48^3 \times 144$ & 4.52 & 62 & 1600 & 5.42 & 168 \\
        \hline
        a09m130 & 0.0871(06) & 0.2871 &138.1(1.0) & 128.2(1) & $64^3 \times 96$ & 3.90 & 40 & 800 & 6.07 & 45 \\
        a09m220 & 0.0872(07) & 0.2873 & 225.9(1.8) & 220.3(2) & $48^3 \times 96$ & 4.79 & 40 & 800 & 6.07 & 93 \\
        a09m310 & 0.0888(08) & 0.2897 & 313.0(2.8) & 312.7(6) & $32^3 \times 96$ & 4.51 & 40 & 800 & 6.31 & 156 \\
        \hline 
        a12m220 & 0.1184(09) & 0.3348 &227.9(1.9) & 216.9(2) & $32^3 \times 64$ & 4.38 & 30 & 400 & 5.61 & 150  \\
        a12m220L & 0.1189(09) & 0.3348 &227.6(1.7) & 217.0(2) & $40^3 \times 64$ & 5.49 & 30 & 400 & 5.65 & 150  \\
        a12m310 & 0.1207(11) & 0.3384 & 310.2(2.8) & 305.3(4) & $24^3 \times 64$ & 4.55 & 30 & 400 & 5.83 & 179  \\
        \hline 
        a15m310 & 0.1510(20) & 0.3881 &320.6(4.3) & 306.9(5) & $16^3 \times 48$ & 3.93 & 24 & 400 & 9.12 & 80  \\
        \hline \hline 
    \end{tabular}
    \end{center}
    \caption{The eight HISQ-Clover lattice ensembles used in this work. To increase statistics, each configuration is divided into 8 sublattice regions and in each we make 256 measurements for diagram A and 64 for diagram C. The values of $a$, $M_\pi^{\rm sea}$ and $M_\pi^{\rm val}$ are reproduced from Ref.~\cite{Gupta:2018qil}.}
    \label{tab:ensembles}
\end{table*}

\section{Error reduction in the extraction of $\mathcal{H}_{\mu \nu}^{V A}$}
\label{sec:err_red}

The spectral decomposition of the two-point correlator of the pion is:
\begin{align}
    C_{2pt}(\tau) &= C^{fwd}_{2pt}(\tau) + C^{bkw}_{2pt}(\tau) \nonumber\\
       &= \sum_{\mathbf{x}} e^{-i\mathbf{p}\cdot \mathbf{x}}\langle J_\pi(\tau,\mathbf{x}) J_\pi^\dag(0) \rangle \nonumber\\
    &= \sum_{i} \Big| \langle 0 | J_\pi | \pi_i (p) \rangle \Big|^2 \frac{e^{-E_i(\mathbf{p})\tau } + e^{-E_i(\mathbf{p})( T-\tau )}}{2E_i(\mathbf{p})} 
\end{align} 
where $i$ indexes the states. Statistics for $C_{2pt}(\tau)$ are  increased by averaging over forward and backward propagation. 
From here on, we will truncate the sum over states to just the ground state contribution since the ($V,A$) insertions can both be made in the plateau region, i.e., far enough away from both the source and the sink time slices where contribution of excited states is negligible. 

The spectral decomposition of the hadronic tensor, limited to zero momentum source and sink by using wall sources for quark propagators,  and normalized by the 2-point function, is
\begin{align}
    {}R^H_{\mu \nu} &(t, \tau, \vec{x}) 
    = \frac{ C_{4pt}(\tau, t, \vec{x})}{C_{2pt}(\tau)}\nonumber\\
    &\buildrel{\scriptstyle \tau \rightarrow 0}\over{=}\frac{ 2M_\pi \langle J_{\pi^0}(\tau/2) J^{em}_\mu(0,0)J^{W,A}_\nu({\vec x},t) J_{\pi^{-}}(-\tau/2)\rangle}{ \Big| \langle 0 | J_\pi | \pi \rangle \Big|^2 e^{-M_\pi\tau } } \nonumber \\ 
    &\buildrel{\scriptstyle \tau \rightarrow 0}\over{=}
    \langle \pi^0(p)|T\left[J^{em}_\mu(0,0)J^{W,A}_\nu({\vec x},t)\right]|\pi^{-}(p)\rangle / 2M_\pi \nonumber\\
    &\buildrel{\phantom{\scriptstyle \tau \rightarrow 0}}\over{=}\mathcal{H}_{\mu \nu} / 2M_\pi
\label{eq:R4to2}
\end{align}
where for $C_{2pt}(\tau)$ one can use the fit or the data. The second line holds for the kaon as well since our calculation is done in the SU(3) symmetry approximation.

\begin{figure}[ht]  
    \centering
    \includegraphics[width=.49\textwidth]{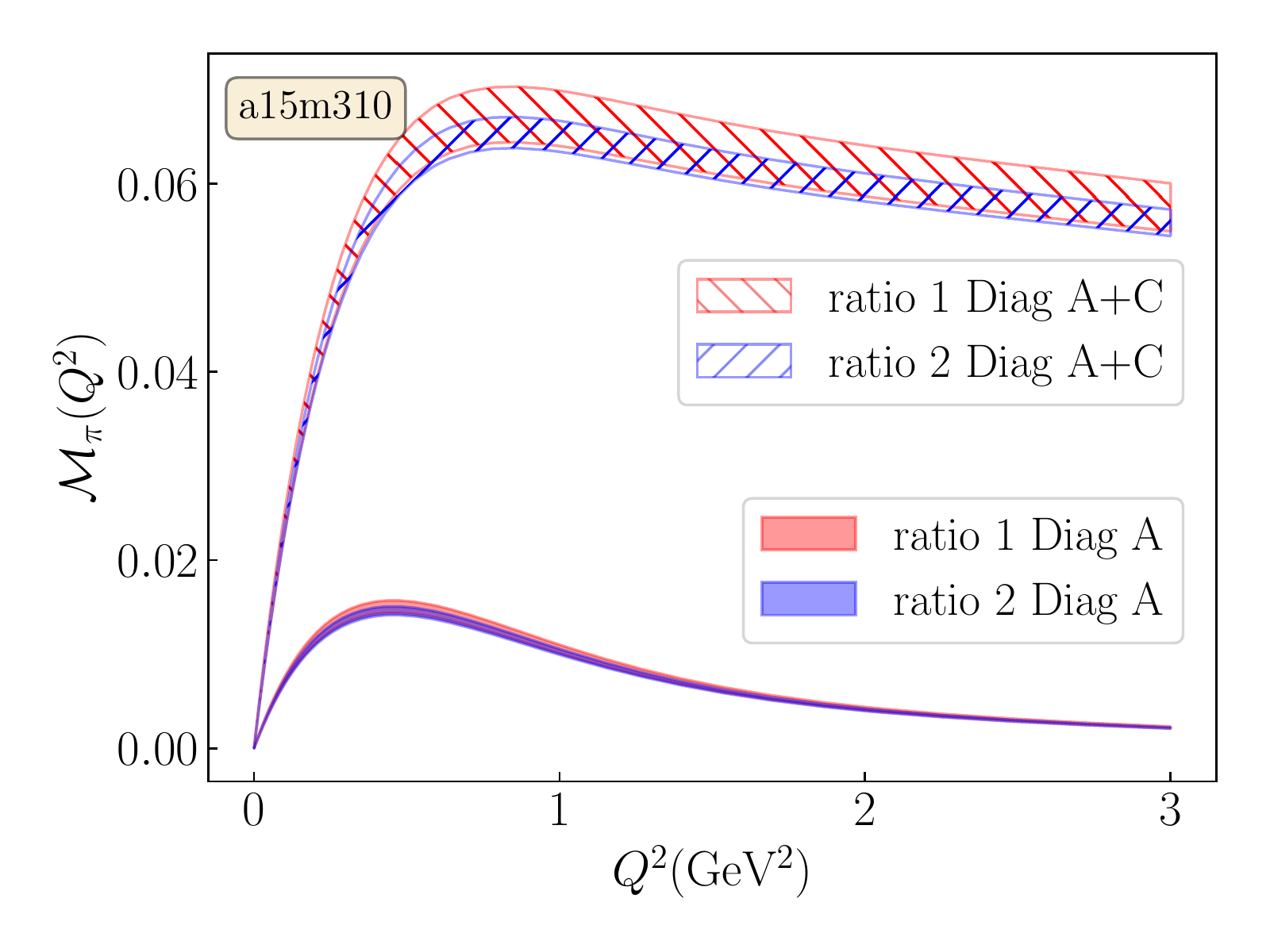}
    \vspace{-0.2cm}
    \caption{Comparison of the signal in ${\cal M}_\pi (Q^2)$ extracted by (i) method 1 combining the ratio defined in Eq.~\protect\eqref{eq:R4to2} and $F_+^{\pi} = \sqrt{2}$ (red), and (ii) method 2 using the ratio in Eq.~\protect\eqref{eq:R4to3} (blue). There is roughly a factor of two reduction in errors using Eq.~\protect\eqref{eq:R4to3} as deduced  by comparing the blue and red bands.\looseness-1}
    \label{fig:Signal_Mpi}
\end{figure}

The form factor $F^H_+ $ (matrix element) is obtained from the 3-point function,
\begin{equation}
    F^H_+ = \frac{\langle H(p') |J^{W,V}_\mu | H(p) \rangle }{(p+p')_{\mu=4}} =\frac{\sqrt{2}C_{3pt}(\tau)}{ C_{2pt}(\tau)},
     \label{eq:R3to2}
\end{equation}
for $H=\pi$. For $H=K$, the factor $\sqrt 2$ is absent.  Thus, we can calculate the desired ratios 
\begin{align}
    \frac{\mathcal{H}_{\mu \nu}^{V A}(t, \vec{x})}{F_{+}^{\pi}} &= 2M_\pi \frac{C_{4pt}(\tau, t, \vec{x})}{\sqrt{2}C_{3pt}(\tau)} \\
    \frac{\mathcal{H}_{\mu \nu}^{V A}(t, \vec{x})}{F_{+}^{K}}
    &= 2M_K \frac{C_{4pt}(\tau, t, \vec{x})}{C_{3pt}(\tau)}
     \label{eq:R4to3}
\end{align}
in two ways: first, using the left hand side with  ($F_+^\pi(0) = \sqrt{2}$ and $ F_+^K(0) = 1$), where the factor $\sqrt{2}$ comes from the normalization of the pion states. In the second method, we use the ratio of correlation functions in the right hand side of Eq.~\eqref{eq:R4to3}. As shown in Fig.~\ref{fig:Signal_Mpi}, the errors in the 3- and 4-point functions are correlated and partially cancel in the second method, which we therefore use for the final results.

\vspace{.5cm}

\section{Comparing lattice results for ${\cal M}_H(Q^2)$ with perturbation theory}
\label{sec:pert}

\begin{figure}  
    \centering
    \includegraphics[width=.48\textwidth]{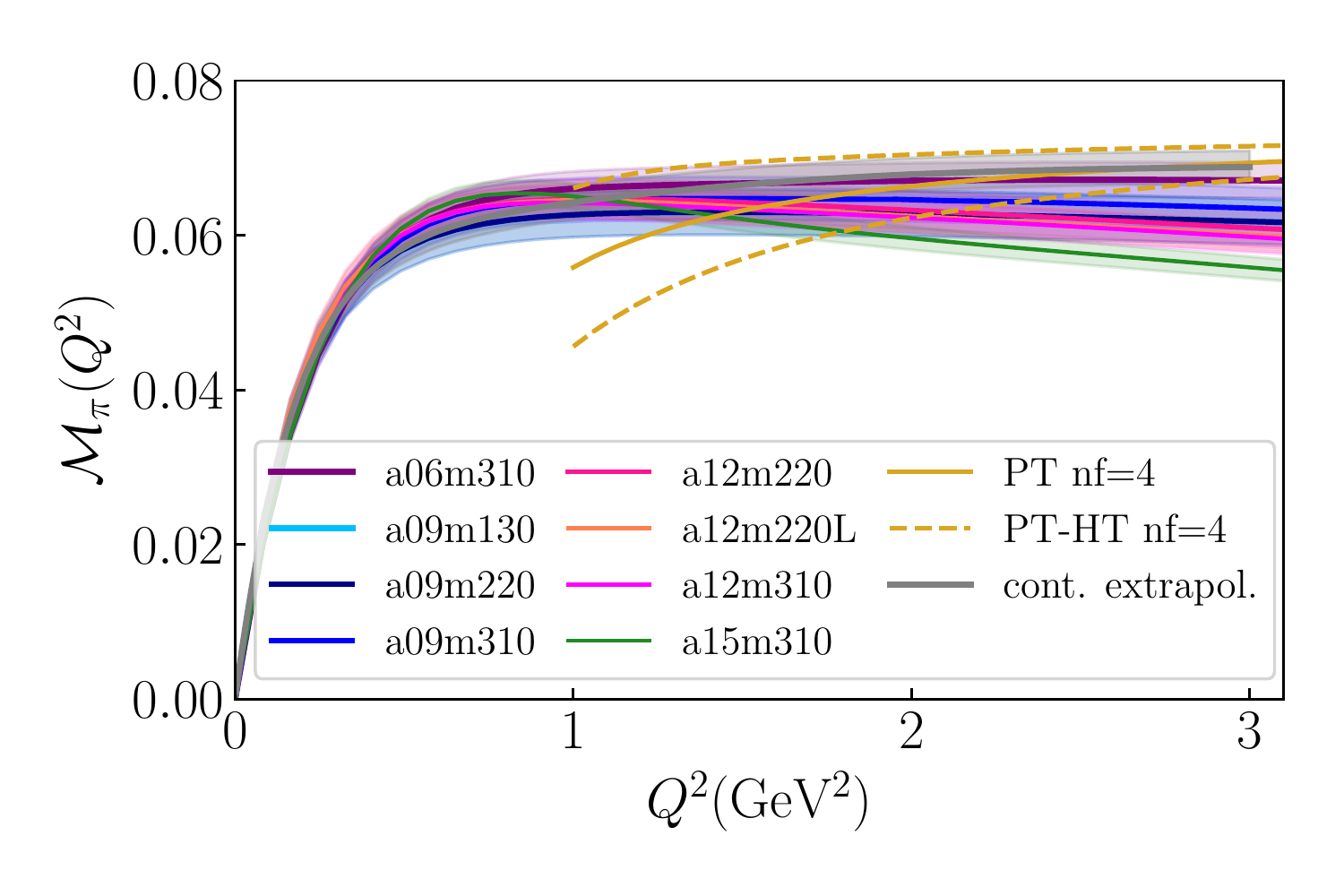}\\
    \vspace{-.5cm}\includegraphics[width=.48\textwidth]{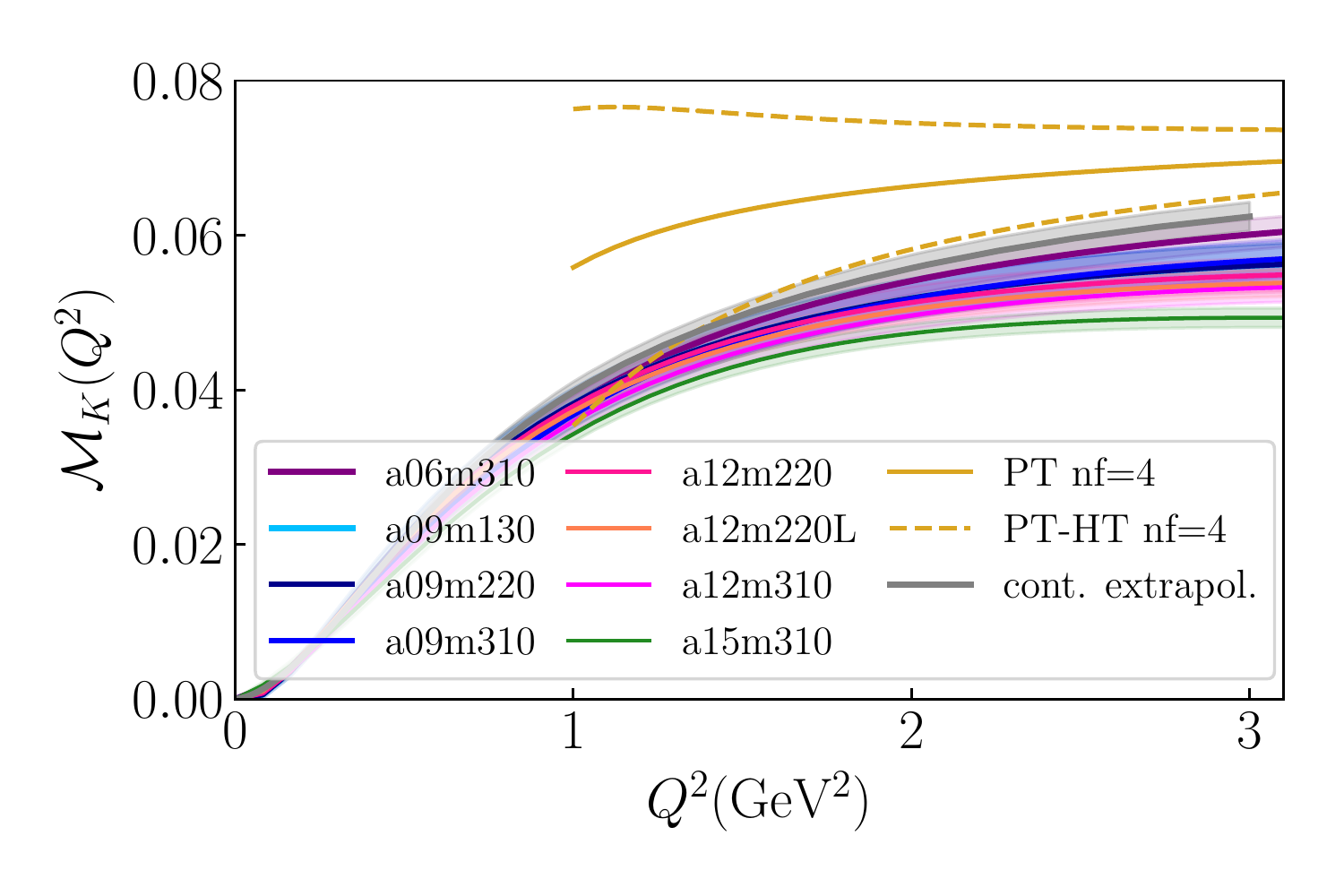}\\
    \vspace{-.47cm}\includegraphics[width=.48\textwidth]{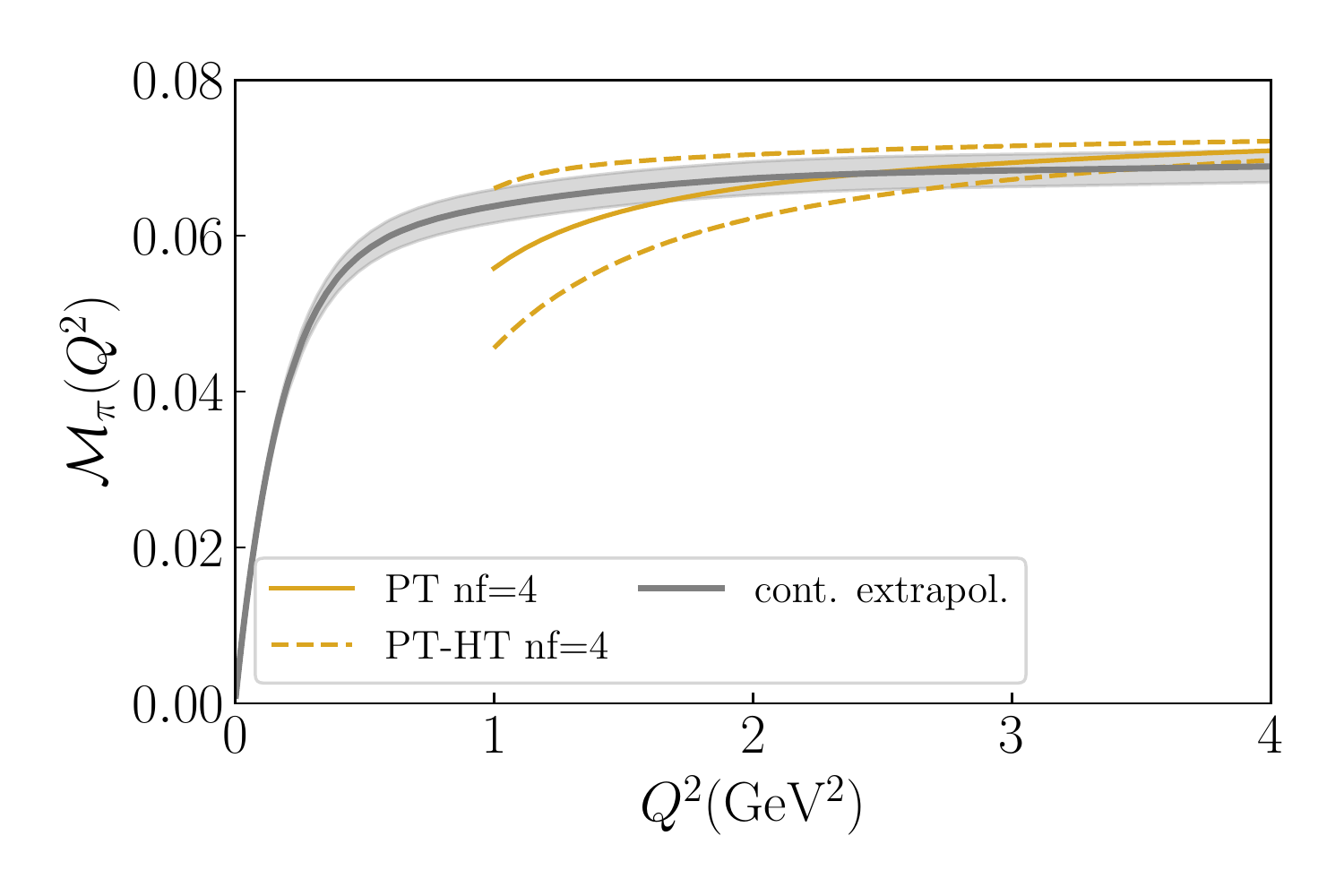}\\
    \vspace{-.47cm}\includegraphics[width=.48\textwidth]{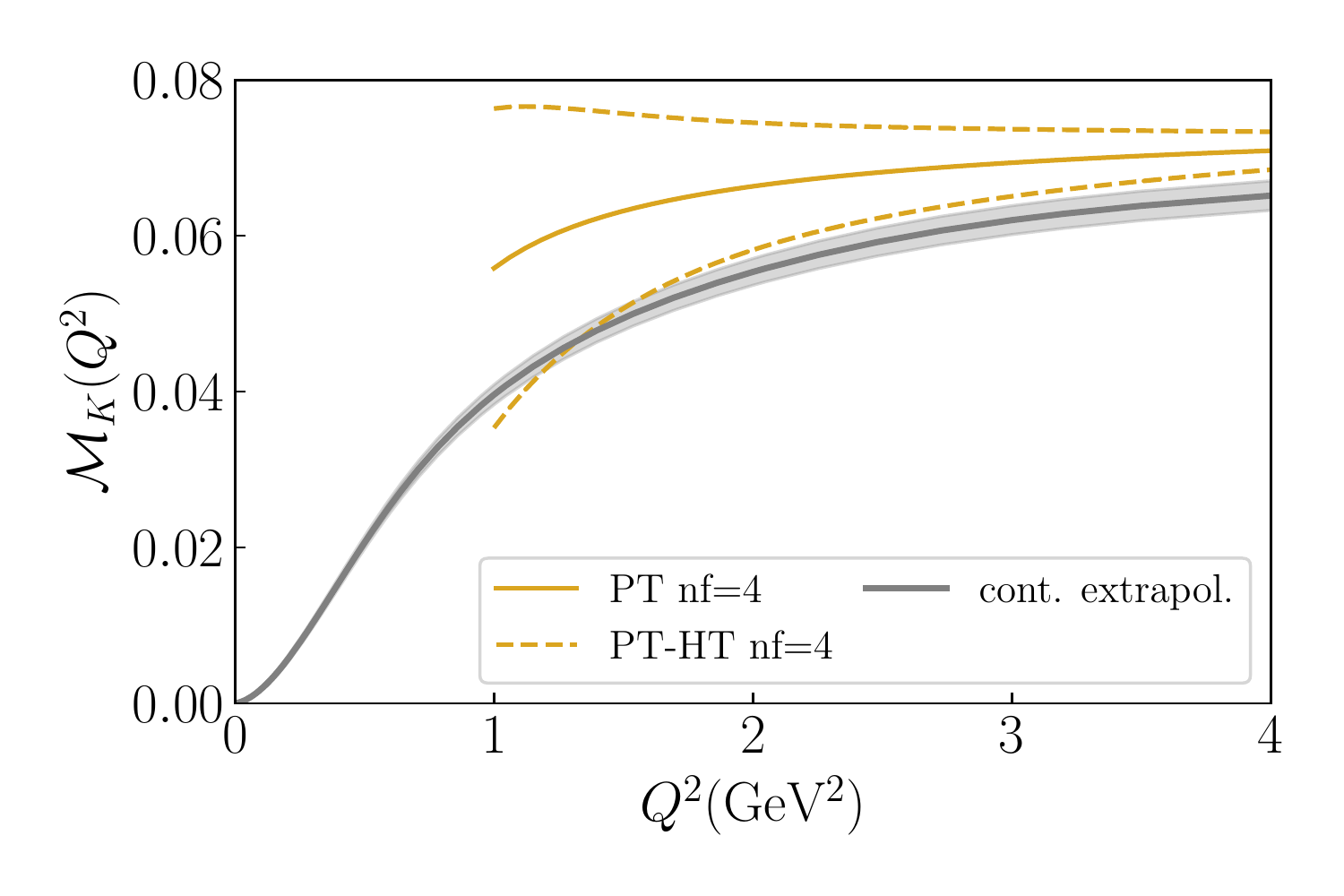}
    \vspace{-.7cm}\caption{$M_H(Q^2)$ for the (a) pion and (b) the kaon for the eight ensembles (top). The bottom panels zoom in on the comparison between the grey band obtained by making a continuum extrapolation at each of the 60 $Q^2$ values and the gold line shows the perturbative result with uncertainty band reflecting higher-twist corrections.\looseness-1}
    \label{fig:M_H}
\end{figure}

\begin{figure*}  
    \centering
    \includegraphics[width=.49\textwidth]{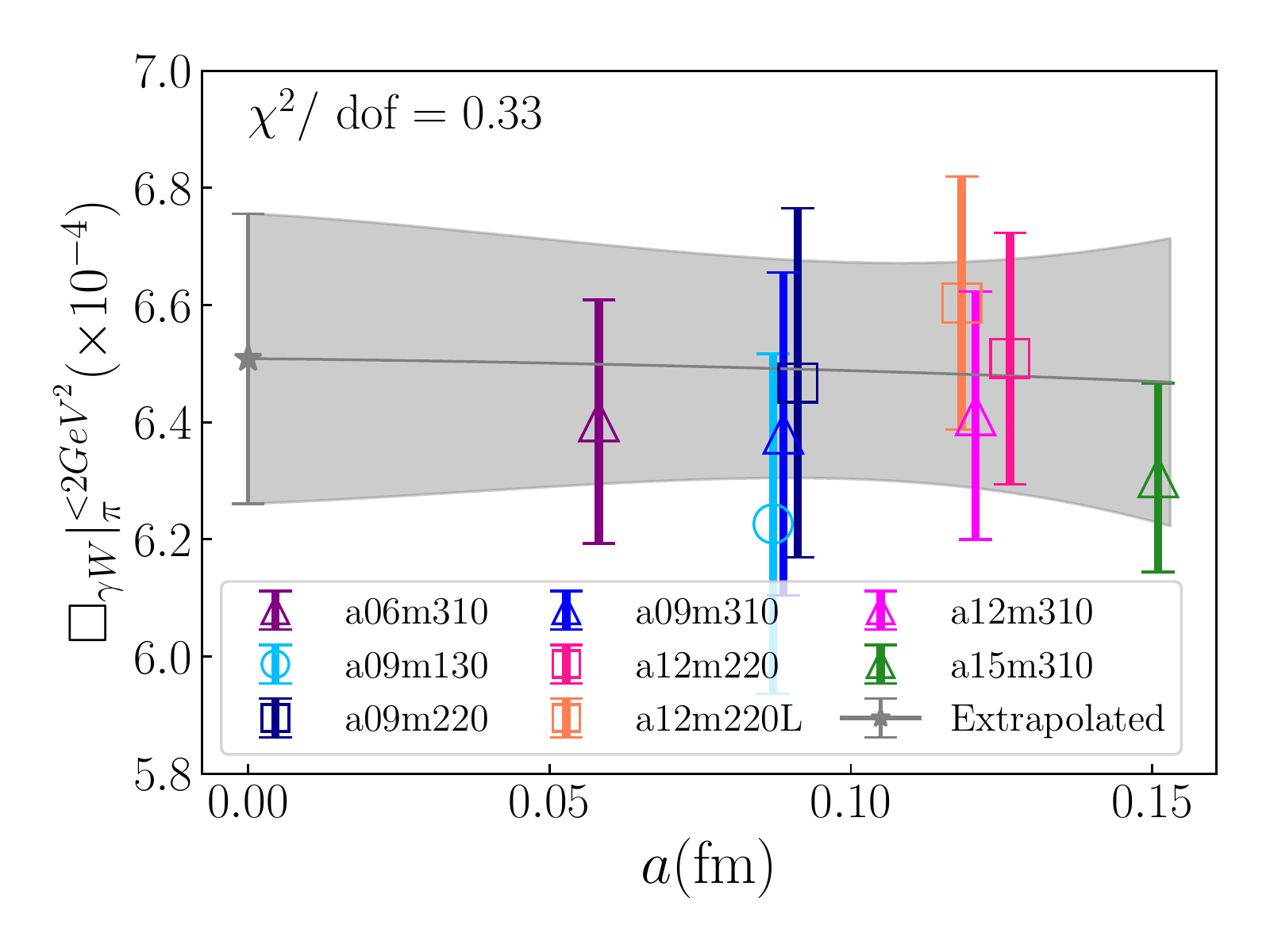}
    \includegraphics[width=.49\textwidth]{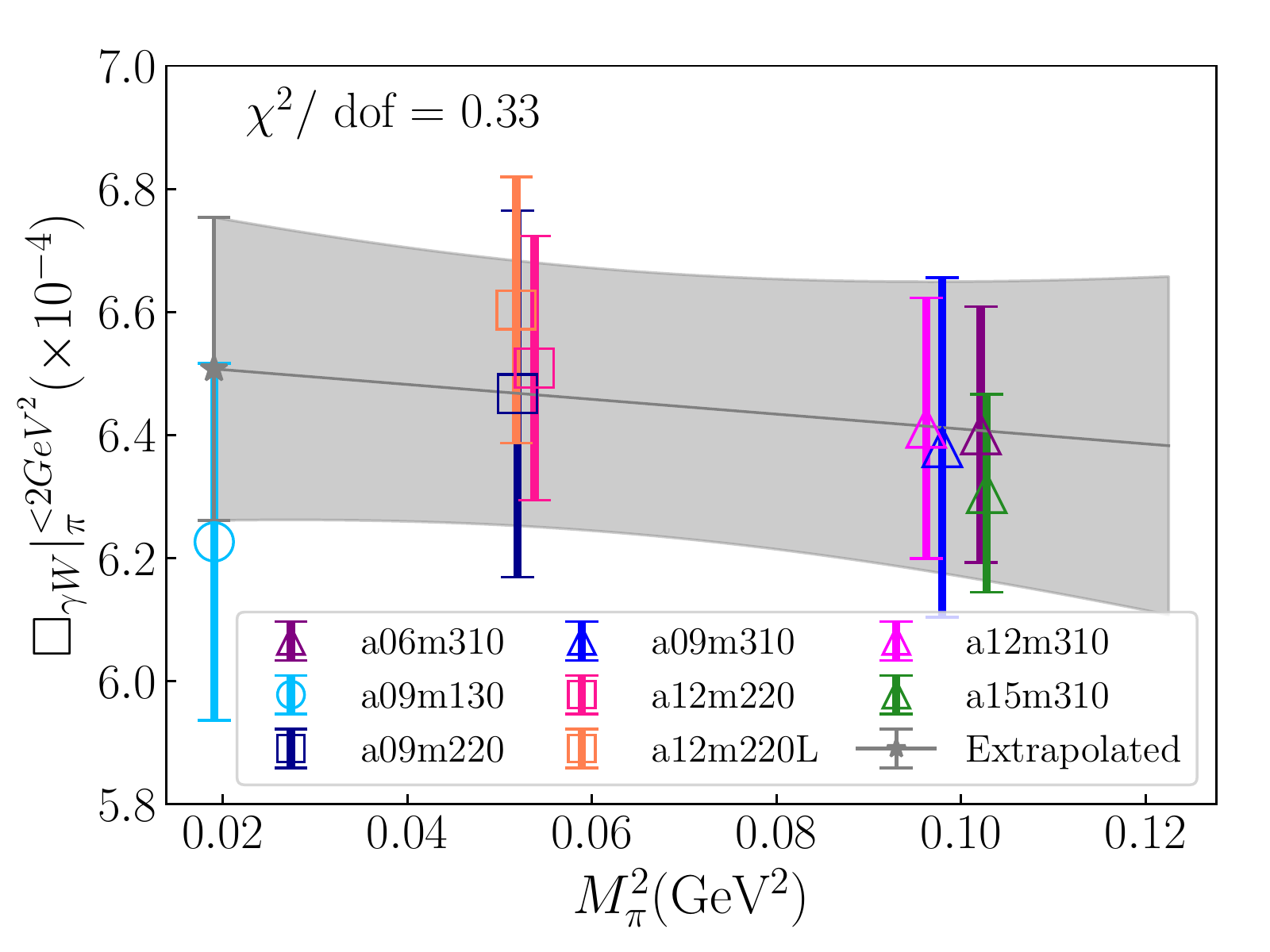}
    \label{fig:cont_chiral_extrapol_pi} \\
    \vspace{-0.2cm}
    \includegraphics[width=.49\textwidth]{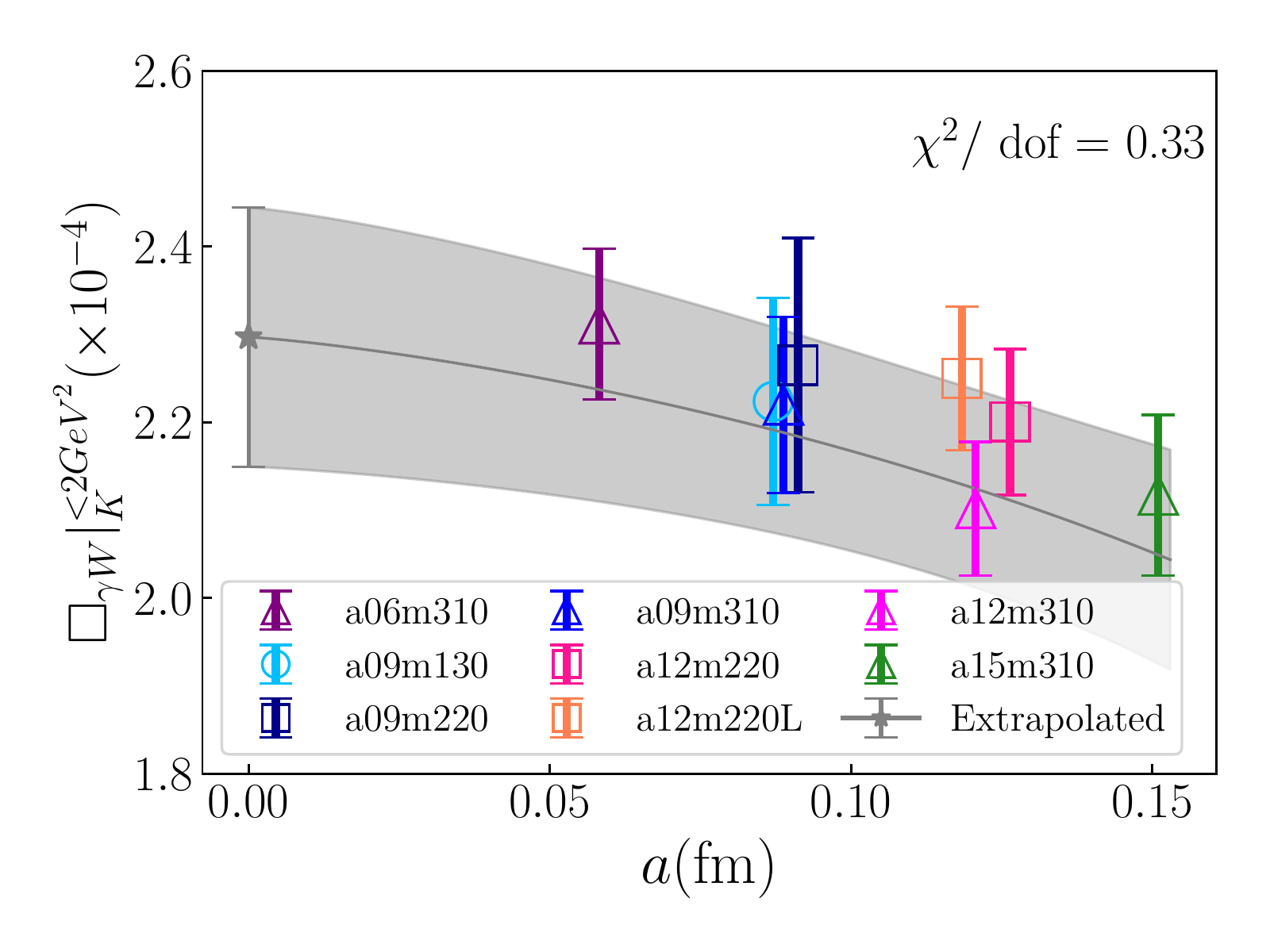}
    \includegraphics[width=.49\textwidth]{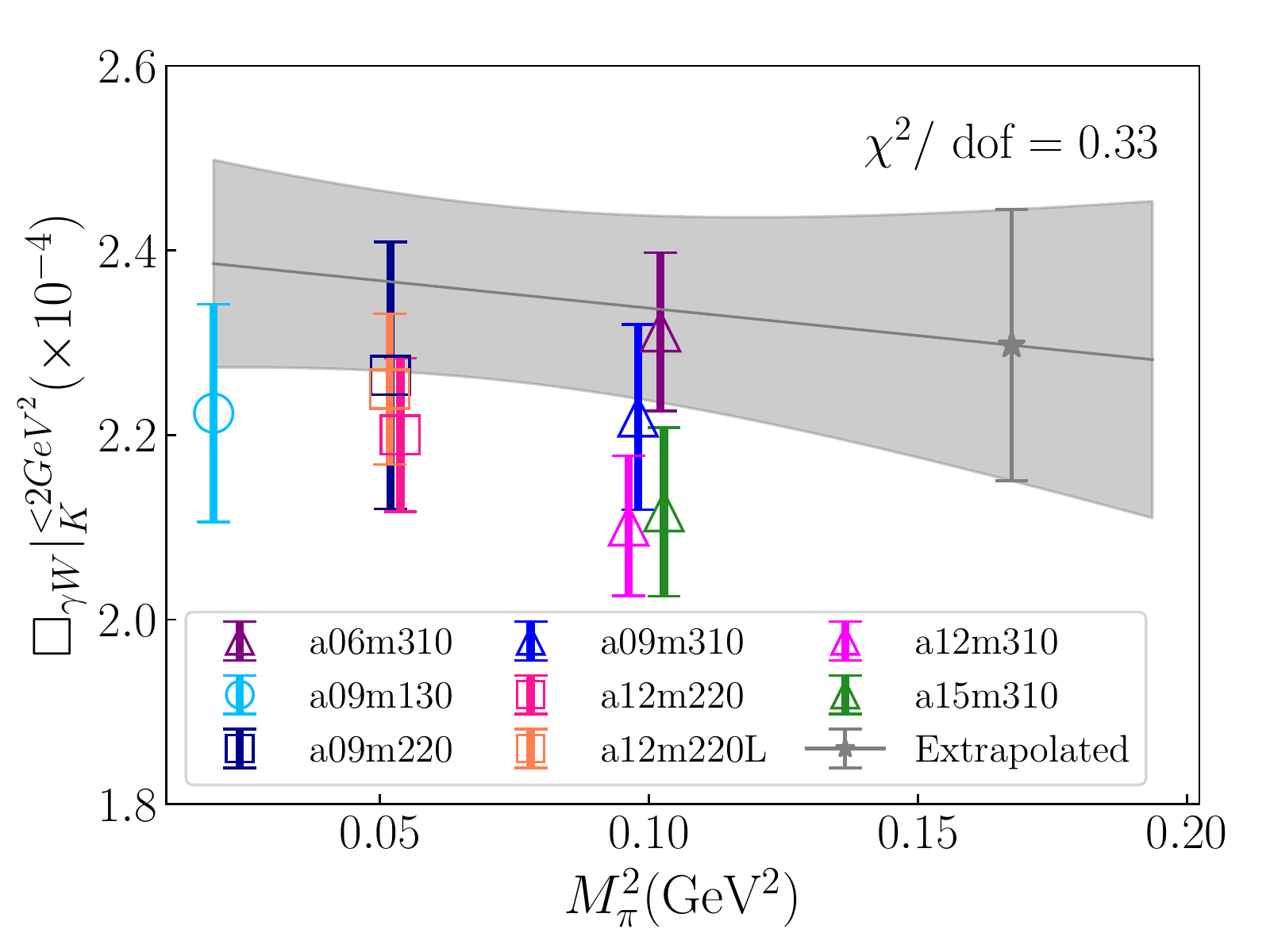}
    \vspace{-0.2cm}
    \caption{The dependence of the $\gamma W$-box contribution for $Q^2 \le 2 \textrm{GeV}^2$ for the pion (top) and kaon (bottom)  on  the lattice spacing $a$ (left), and  the pion mass ($M_\pi^2$) (right). The symbols for the various ensembles are defined in the inset and in Table~\protect\ref{tab:ensembles}. The results at the physical point are 
    shown by the grey star symbol. The result for the kaon is evaluated at the SU(3) symmetric point.}
    \label{fig:cont_chiral_extrapol_pi_K}
\end{figure*}

As mentioned in Sec.~\ref{sec:EWbox},  ${\cal M}_H$ can be extracted
at all values of $Q^2$. In practice, we choose sixty $Q^2$ values that are the same on all eight ensembles with a higher density below $Q^2 < 1$~GeV${}^2$.  These 60 points are converted into the smooth curves shown in Fig.~\ref{fig:M_H} (top) using the cubic spline interpolator from scipy library~\cite{2020SciPy-NMeth}. Data show that as $Q^2$ 
increases above $1$~GeV${}^2$, the value of ${\cal M}_H $ on coarser lattices 
decreases, indicating a dependence on the lattice spacing. Below $Q^2 < 1$~GeV${}^2$, 
the trend reverses. The integrated box contributions for $Q^2 < 2$~GeV${}^2$ and their dependence on $a$ and $M_\pi^2$ is shown in Fig.~\ref{fig:cont_chiral_extrapol_pi_K}. \looseness-1

\begin{figure*}  
    \centering
    \includegraphics[width=.49\textwidth]{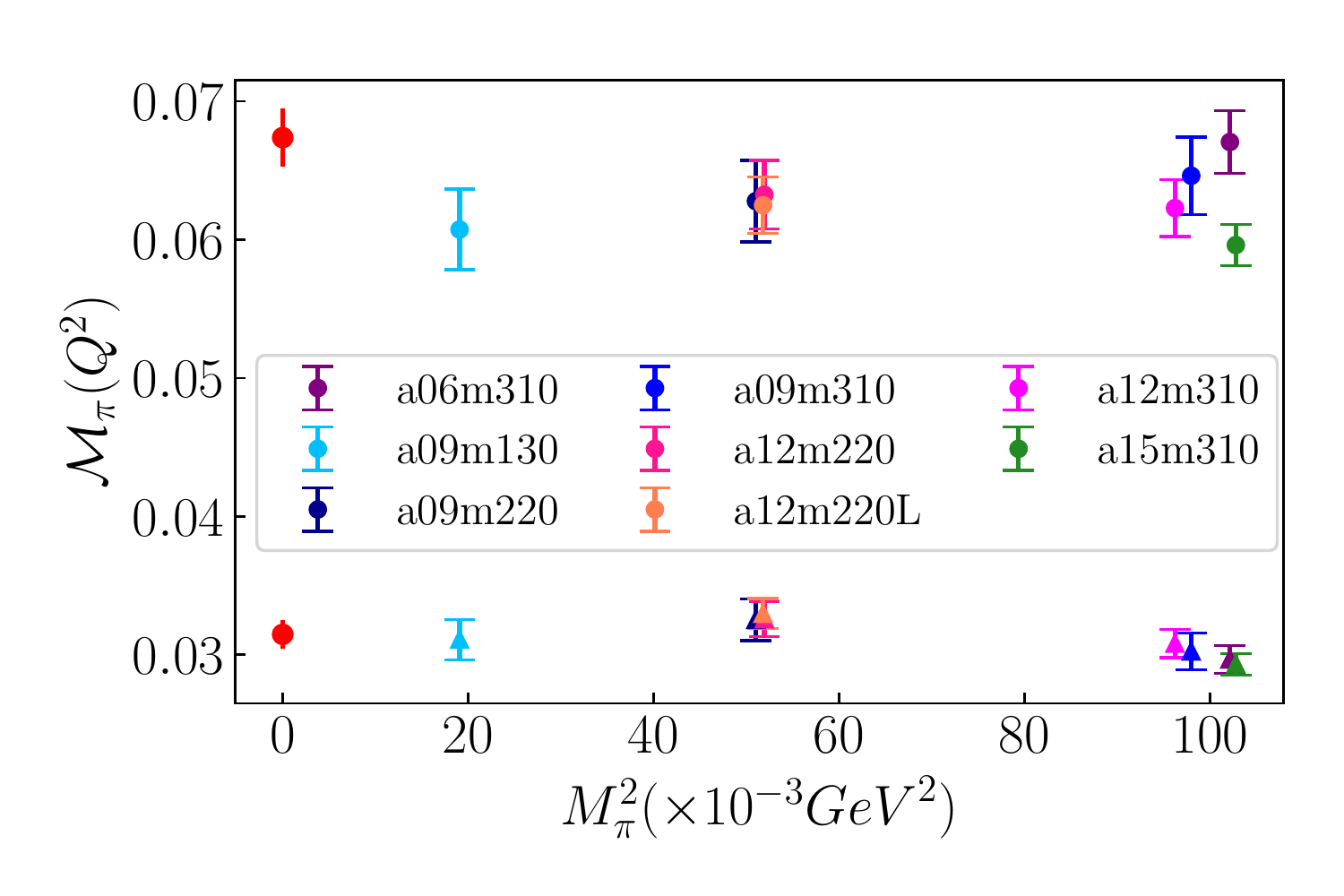}
    \includegraphics[width=.49\textwidth]{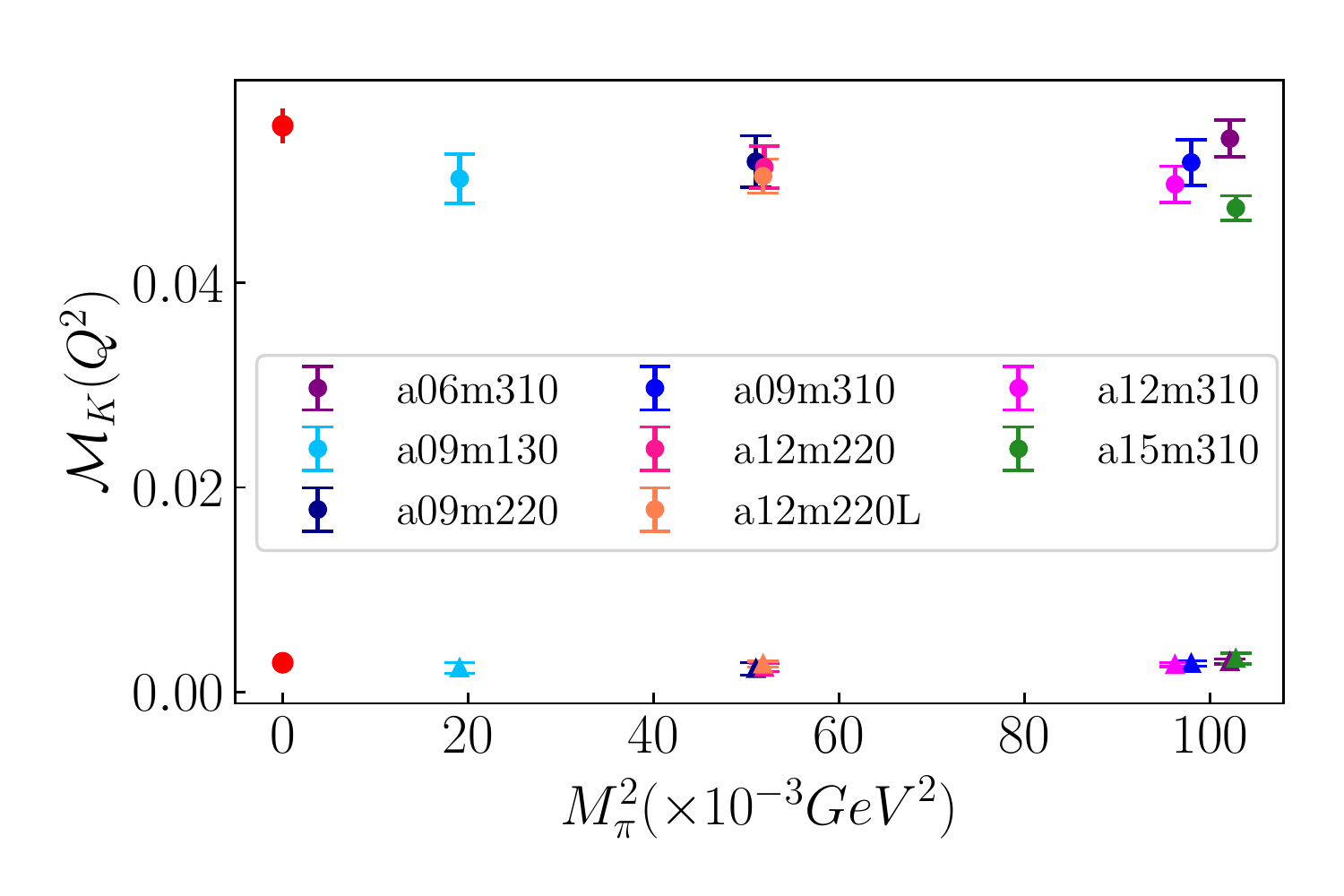}
     \vspace{-0.2cm}
    \caption{${\cal M}_H(Q^2)$ for the pion (left) and kaon (right) at $Q^2 = 0.133\textrm{ GeV}^2$ (triangles), $2.00\ \textrm{GeV}^2$ (circles). Ensembles are labeled by color. Data for ${\cal M}_H (Q^2) $ do not show a significant dependence on $M_\pi^2$. The red points on the very left in each panel are the continuum extrapolated values using a fit linear in $a\alpha_S$, i.e., ignoring possible dependence on $M_\pi^2$. }
    \label{fig:mass_dep}
\end{figure*}

To compare the lattice ${\cal M}_H(Q^2)$ to perturbation theory, we extrapolate the data to the continuum limit at $M_\pi = 135$~MeV using a fit linear in just $\alpha_S\, a$ since the dependence on $M_\pi$ is observed to be small (See Fig.~\ref{fig:mass_dep}). These fits, for all the ensembles and all $Q^2$ values, have a $p$-value above 0.2. 
As shown in Figure~\ref{fig:M_H}, this continuum limit data,   represented by the grey solid line, roughly agrees with perturbative result using the operator product expansion~\cite{Larin:1991tj,Baikov:2010je,Feng:2020zdc} (gold line) for $Q^2 >  2 \textrm{GeV}^2$. Uncertainty in the perturbative result arises from the truncation 
of the  series at the \(4^{\rm th}\) order and the neglected higher-twist (HT) contributions~\cite{Feng:2020zdc}. Since diagram (A) only has HT contributions, we use its lattice value as an estimate of the HT uncertainty and show this by the dotted lines about the perturbative result.

  \vspace{-0.8cm}

\section{Continuum extrapolation of the lattice data}
\label{sec:CCFV}

The extrapolation of the $\gamma W$-box for $Q^2 < Q_{\text{cut}}^2  = 2$~GeV${}^2$ to the continuum limit $a=0$ and pseudoscalar mass $M_\pi = M_\pi^{\text{phys}}$ for the pion, and $M_\pi = M_K^{\text{SU(3)}}$ for the kaon is carried out keeping the lowest order dependence on the pion mass ($M_\pi^2$) and on the lattice spacing ($\alpha_S\, a$):
\begin{equation}
    \Box |_{VA}^{Q^2 < Q_{\text{cut}}^2} ( M_\pi, a) = c_0 + c_1 \alpha_S\,a + c_2 M_\pi^2 \,.
\end{equation}
This extrapolation is shown in (Fig.~\ref{fig:cont_chiral_extrapol_pi_K}) and gives
    \begin{align}
        \square_{\gamma W}^{VA} |^{Q^2 \le 2 \textrm{GeV}^2}_{\pi}
        &= 0.651 (25) \times 10^{-3}   \,,\\
        \square_{\gamma W}^{VA} |^{Q^2 \le 2 \textrm{GeV}^2}_{K}
        &= 0.230 (15) \times 10^{-3} \,, 
        \label{eq:box_pi_K}
    \end{align}
Systematic uncertainties due to the chiral-continuum extrapolation are included in these estimates. 
We also explored possible uncertainty in ${\cal M}_H$ due to approximating the integral 
over $Q^2 \le 2$~GeV${}^2$  using 
52 discrete points by comparing the trapezoid and Simpson methods and found 
the difference to be negligible. 
 We also assume that finite volume effects are negligible  since all ensembles have $M_\pi L \ge 3.9$.

\section{Results for the $\gamma W$-box diagram and comparison to earlier works}
\label{sec:results}

The contribution above the energy cut at $Q^2 = 2 \,\textrm{GeV}^2$ is computed using the operator product expansion~\cite{Feng:2020zdc} with the higher-twist uncertainty estimated using diagram A (See Fig.~\ref{fig:diagrams}).
    \begin{equation}
    \label{eq:PT}
        \square_{\gamma W}^{VA}|_{\pi,K}^{Q^2 > 2 \textrm{GeV}^2} = 2.159(6)_{HO}(7)_{HT}\times 10^{-3}.
    \end{equation}
Combining Eq.~\eqref{eq:PT} with  Eq.~\eqref{eq:box_pi_K} gives our results for the full box contribution:
    \begin{align}
        \square_{\gamma W}^{VA} |_{\pi}
        &= 2.810 (26) \times 10^{-3} \,, \\
        \square_{\gamma W}^{VA} \Big|_{K^{0, S U(3)}}
        &= 2.389 (17) \times 10^{-3} \,,
    \end{align}
which are in good agreement with those from Feng et al.~\cite{Feng:2020zdc,Ma:2021azh}
    \begin{align}
         \square_{\gamma W}^{VA}|_{\pi} &= 2.830(11)(26) \times 10^{-3} \,, \\
        \left.\square_{\gamma W}^{V A}\right|_{K^{0, S U(3)}} &= 2.437(44) \times 10^{-3} \,.
    \end{align}
    The difference in $\square_{\gamma W}^{V A}\Big|_{K^{0, S U(3)}}$ is $1.09 \sigma$, but note that our value is determined with extrapolation in $M_\pi^2$ to $SU(3)$-symmetric point, while the Feng et al.\ value, also called $\square_{\gamma W}^{V A}|_{K^{0, S U(3)}}$, was computed at the physical pion mass in all five ensembles, i.e., without extrapolation to $M_K|_{\rm SU(3)}$.

    The agreement between the two calculations provides important consistency checks as they are done at different values of $\{a,M_\pi\}$ (see Fig.~\ref{fig:latt_param}) and with different lattice actions. The largest uncertainty in the results presented in~\cite{Feng:2020zdc,Ma:2021azh} comes from the systematic difference  between DSDR and Iwasaki estimates, whereas in our calculation it comes from the renormalization constant $Z_A$ as shown in Fig.~\ref{fig:error_budget}, which is unity for domain-wall fermions. 
    
    Our data from eight ensembles with the same action provides a more controlled 
    chiral-continuum extrapolation than in~\cite{Feng:2020zdc,Ma:2021azh}. The data for the pion display no significant dependence on $a$ or $M_\pi^2$. The data for the kaon in Fig.~\ref{fig:cont_chiral_extrapol_pi_K} shows $\approx 10\%$ dependence on $a$ but is again flat with respect to $M_\pi^2$. A similar level of dependence on $a$ was found in the Iwasaki action data in Ref.~\cite{Ma:2021azh}.

    To conclude, taking the two calculations together, increases our confidence that calculations of the $\gamma W$-box part of the radiative corrections to pion and kaon decays using lattice QCD are robust. The analysis of RC to neutron decays is, as expected, turning out to be much more challenging because of the more severe signal-to-noise and contamination from excited states problems.  
    
\vspace{0.2cm}
\noindent {\textbf {Acknowledgments}}: We thank the MILC collaboration for providing the HISQ lattices, and Vincenzo Cirigliano and Emanuele Mereghetti for discussions. The calculations used the CHROMA software suite~\cite{Edwards:2004sx}. Simulations were carried out at (i) the NERSC supported by DOE under Contract No. DE-AC02-05CH11231;  (ii) the Oak Ridge Leadership Computing Facility, which is a DOE Office of Science User Facility supported under Award No. DE-AC05-00OR22725 through the INCITE program project HEP133, (iii) the USQCD collaboration resources funded by DOE HEP, and (iv) Institutional Computing at Los Alamos National Laboratory. This work was supported by LANL LDRD program and TB and RG were also supported by the DOE HEP  under Contract No. DE-AC52-06NA25396.
\bibliographystyle{apsrev4-2} 
\showtitleinbib
\bibliography{ref}

\providecommand{\noopsort}[1]{}\providecommand{\singleletter}[1]{#1}%
\begin{thebibliography}{27}%
\makeatletter
\providecommand \@ifxundefined [1]{%
 \@ifx{#1\undefined}
}%
\providecommand \@ifnum [1]{%
 \ifnum #1\expandafter \@firstoftwo
 \else \expandafter \@secondoftwo
 \fi
}%
\providecommand \@ifx [1]{%
 \ifx #1\expandafter \@firstoftwo
 \else \expandafter \@secondoftwo
 \fi
}%
\providecommand \natexlab [1]{#1}%
\providecommand \enquote  [1]{``#1''}%
\providecommand \bibnamefont  [1]{#1}%
\providecommand \bibfnamefont [1]{#1}%
\providecommand \citenamefont [1]{#1}%
\providecommand \href@noop [0]{\@secondoftwo}%
\providecommand \href [0]{\begingroup \@sanitize@url \@href}%
\providecommand \@href[1]{\@@startlink{#1}\@@href}%
\providecommand \@@href[1]{\endgroup#1\@@endlink}%
\providecommand \@sanitize@url [0]{\catcode `\\12\catcode `\$12\catcode
  `\&12\catcode `\#12\catcode `\^12\catcode `\_12\catcode `\%12\relax}%
\providecommand \@@startlink[1]{}%
\providecommand \@@endlink[0]{}%
\providecommand \url  [0]{\begingroup\@sanitize@url \@url }%
\providecommand \@url [1]{\endgroup\@href {#1}{\urlprefix }}%
\providecommand \urlprefix  [0]{URL }%
\providecommand \Eprint [0]{\href }%
\providecommand \doibase [0]{https://doi.org/}%
\providecommand \selectlanguage [0]{\@gobble}%
\providecommand \bibinfo  [0]{\@secondoftwo}%
\providecommand \bibfield  [0]{\@secondoftwo}%
\providecommand \translation [1]{[#1]}%
\providecommand \BibitemOpen [0]{}%
\providecommand \bibitemStop [0]{}%
\providecommand \bibitemNoStop [0]{.\EOS\space}%
\providecommand \EOS [0]{\spacefactor3000\relax}%
\providecommand \BibitemShut  [1]{\csname bibitem#1\endcsname}%
\let\auto@bib@innerbib\@empty
\bibitem [{\citenamefont {Aoki}\ \emph
  {et~al.}(2022{\natexlab{a}})\citenamefont {Aoki} \emph
  {et~al.}}]{Aoki:2021kgd}%
  \BibitemOpen
  \bibfield  {author} {\bibinfo {author} {\bibfnamefont {Y.}~\bibnamefont
  {Aoki}} \emph {et~al.} (\bibinfo {collaboration} {Flavour Lattice Averaging
  Group (FLAG)}),\ }\bibfield  {title} {\bibinfo {title} {{FLAG Review 2021}},\
  }\href {https://doi.org/10.1140/epjc/s10052-022-10536-1} {\bibfield
  {journal} {\bibinfo  {journal} {Eur.\ Phys.\ J.}\ }\textbf {\bibinfo {volume}
  {C82}},\ \bibinfo {pages} {869} (\bibinfo {year} {2022}{\natexlab{a}})},\
  \Eprint {https://arxiv.org/abs/2111.09849} {arXiv:2111.09849 [hep-lat]}
  \BibitemShut {NoStop}%
\bibitem [{\citenamefont {Workman}\ and\ \citenamefont
  {Others}(2022)}]{Workman:2022ynf}%
  \BibitemOpen
  \bibfield  {author} {\bibinfo {author} {\bibfnamefont {R.~L.}\ \bibnamefont
  {Workman}}\ and\ \bibinfo {author} {\bibnamefont {Others}} (\bibinfo
  {collaboration} {Particle Data Group}),\ }\bibfield  {title} {\bibinfo
  {title} {{Review of Particle Physics}},\ }\href
  {https://doi.org/10.1093/ptep/ptac097} {\bibfield  {journal} {\bibinfo
  {journal} {PTEP}\ }\textbf {\bibinfo {volume} {2022}},\ \bibinfo {pages}
  {083C01} (\bibinfo {year} {2022})}\BibitemShut {NoStop}%
\bibitem [{\citenamefont {Gonzalez}\ \emph {et~al.}(2021)\citenamefont
  {Gonzalez} \emph {et~al.}}]{UCNt:2021pcg}%
  \BibitemOpen
  \bibfield  {author} {\bibinfo {author} {\bibfnamefont {F.~M.}\ \bibnamefont
  {Gonzalez}} \emph {et~al.} (\bibinfo {collaboration}
  {UCN\ensuremath{\tau}}),\ }\bibfield  {title} {\bibinfo {title} {{Improved
  neutron lifetime measurement with UCN$\tau$}},\ }\href
  {https://doi.org/10.1103/PhysRevLett.127.162501} {\bibfield  {journal}
  {\bibinfo  {journal} {Phys.\ Rev.\ Lett.}\ }\textbf {\bibinfo {volume}
  {127}},\ \bibinfo {pages} {162501} (\bibinfo {year} {2021})},\ \Eprint
  {https://arxiv.org/abs/2106.10375} {arXiv:2106.10375 [nucl-ex]} \BibitemShut
  {NoStop}%
\bibitem [{\citenamefont {Hardy}\ and\ \citenamefont
  {Towner}(2012)}]{Hardy_2012}%
  \BibitemOpen
  \bibfield  {author} {\bibinfo {author} {\bibfnamefont {J.~C.}\ \bibnamefont
  {Hardy}}\ and\ \bibinfo {author} {\bibfnamefont {I.~S.}\ \bibnamefont
  {Towner}},\ }\bibfield  {title} {\bibinfo {title} {Superallowed \(0+ \to 0+\)
  beta-decay from {Tz} \({} = -1\) sd-shell nuclei},\ }\href
  {https://doi.org/10.1088/1742-6596/387/1/012006} {\bibfield  {journal}
  {\bibinfo  {journal} {Journal of Physics: Conference Series}\ }\textbf
  {\bibinfo {volume} {387}},\ \bibinfo {pages} {012006} (\bibinfo {year}
  {2012})}\BibitemShut {NoStop}%
\bibitem [{\citenamefont {Seng}\ \emph {et~al.}(2018)\citenamefont {Seng},
  \citenamefont {Gorchtein}, \citenamefont {Patel},\ and\ \citenamefont
  {Ramsey-Musolf}}]{Seng:2018yzq}%
  \BibitemOpen
  \bibfield  {author} {\bibinfo {author} {\bibfnamefont {C.-Y.}\ \bibnamefont
  {Seng}}, \bibinfo {author} {\bibfnamefont {M.}~\bibnamefont {Gorchtein}},
  \bibinfo {author} {\bibfnamefont {H.~H.}\ \bibnamefont {Patel}},\ and\
  \bibinfo {author} {\bibfnamefont {M.~J.}\ \bibnamefont {Ramsey-Musolf}},\
  }\bibfield  {title} {\bibinfo {title} {{Reduced Hadronic Uncertainty in the
  Determination of $V_{ud}$}},\ }\href
  {https://doi.org/10.1103/PhysRevLett.121.241804} {\bibfield  {journal}
  {\bibinfo  {journal} {Phys.\ Rev.\ Lett.}\ }\textbf {\bibinfo {volume}
  {121}},\ \bibinfo {pages} {241804} (\bibinfo {year} {2018})},\ \Eprint
  {https://arxiv.org/abs/1807.10197} {arXiv:1807.10197 [hep-ph]} \BibitemShut
  {NoStop}%
\bibitem [{\citenamefont {Seng}\ \emph
  {et~al.}(2019{\natexlab{a}})\citenamefont {Seng}, \citenamefont {Gorchtein},\
  and\ \citenamefont {Ramsey-Musolf}}]{Seng:2018qru}%
  \BibitemOpen
  \bibfield  {author} {\bibinfo {author} {\bibfnamefont {C.~Y.}\ \bibnamefont
  {Seng}}, \bibinfo {author} {\bibfnamefont {M.}~\bibnamefont {Gorchtein}},\
  and\ \bibinfo {author} {\bibfnamefont {M.~J.}\ \bibnamefont
  {Ramsey-Musolf}},\ }\bibfield  {title} {\bibinfo {title} {{Dispersive
  evaluation of the inner radiative correction in neutron and nuclear $\beta$
  decay}},\ }\href {https://doi.org/10.1103/PhysRevD.100.013001} {\bibfield
  {journal} {\bibinfo  {journal} {Phys.\ Rev.}\ }\textbf {\bibinfo {volume}
  {D100}},\ \bibinfo {pages} {013001} (\bibinfo {year} {2019}{\natexlab{a}})},\
  \Eprint {https://arxiv.org/abs/1812.03352} {arXiv:1812.03352 [nucl-th]}
  \BibitemShut {NoStop}%
\bibitem [{\citenamefont {Czarnecki}\ \emph {et~al.}(2019)\citenamefont
  {Czarnecki}, \citenamefont {Marciano},\ and\ \citenamefont
  {Sirlin}}]{Czarnecki:2019mwq}%
  \BibitemOpen
  \bibfield  {author} {\bibinfo {author} {\bibfnamefont {A.}~\bibnamefont
  {Czarnecki}}, \bibinfo {author} {\bibfnamefont {W.~J.}\ \bibnamefont
  {Marciano}},\ and\ \bibinfo {author} {\bibfnamefont {A.}~\bibnamefont
  {Sirlin}},\ }\bibfield  {title} {\bibinfo {title} {{Radiative Corrections to
  Neutron and Nuclear Beta Decays Revisited}},\ }\href
  {https://doi.org/10.1103/PhysRevD.100.073008} {\bibfield  {journal} {\bibinfo
   {journal} {Phys.\ Rev.}\ }\textbf {\bibinfo {volume} {D100}},\ \bibinfo
  {pages} {073008} (\bibinfo {year} {2019})},\ \Eprint
  {https://arxiv.org/abs/1907.06737} {arXiv:1907.06737 [hep-ph]} \BibitemShut
  {NoStop}%
\bibitem [{\citenamefont {Sirlin}(1978)}]{Sirlin:1977sv}%
  \BibitemOpen
  \bibfield  {author} {\bibinfo {author} {\bibfnamefont {A.}~\bibnamefont
  {Sirlin}},\ }\bibfield  {title} {\bibinfo {title} {{Current Algebra
  Formulation of Radiative Corrections in Gauge Theories and the Universality
  of the Weak Interactions}},\ }\href
  {https://doi.org/10.1103/RevModPhys.50.573} {\bibfield  {journal} {\bibinfo
  {journal} {Rev.\ Mod.\ Phys.}\ }\textbf {\bibinfo {volume} {50}},\ \bibinfo
  {pages} {573} (\bibinfo {year} {1978})},\ \bibinfo {note} {[Erratum: Rev.\
  Mod.\ Phys.\ {\bf 50}, 905 (1978)]}\BibitemShut {NoStop}%
\bibitem [{\citenamefont {Feng}\ \emph {et~al.}(2020)\citenamefont {Feng},
  \citenamefont {Gorchtein}, \citenamefont {Jin}, \citenamefont {Ma},\ and\
  \citenamefont {Seng}}]{Feng:2020zdc}%
  \BibitemOpen
  \bibfield  {author} {\bibinfo {author} {\bibfnamefont {X.}~\bibnamefont
  {Feng}}, \bibinfo {author} {\bibfnamefont {M.}~\bibnamefont {Gorchtein}},
  \bibinfo {author} {\bibfnamefont {L.-C.}\ \bibnamefont {Jin}}, \bibinfo
  {author} {\bibfnamefont {P.-X.}\ \bibnamefont {Ma}},\ and\ \bibinfo {author}
  {\bibfnamefont {C.-Y.}\ \bibnamefont {Seng}},\ }\bibfield  {title} {\bibinfo
  {title} {{First-principles calculation of electroweak box diagrams from
  lattice QCD}},\ }\href {https://doi.org/10.1103/PhysRevLett.124.192002}
  {\bibfield  {journal} {\bibinfo  {journal} {Phys.\ Rev.\ Lett.}\ }\textbf
  {\bibinfo {volume} {124}},\ \bibinfo {pages} {192002} (\bibinfo {year}
  {2020})},\ \Eprint {https://arxiv.org/abs/2003.09798} {arXiv:2003.09798
  [hep-lat]} \BibitemShut {NoStop}%
\bibitem [{\citenamefont {Ma}\ \emph {et~al.}(2021)\citenamefont {Ma},
  \citenamefont {Feng}, \citenamefont {Gorchtein}, \citenamefont {Jin},\ and\
  \citenamefont {Seng}}]{Ma:2021azh}%
  \BibitemOpen
  \bibfield  {author} {\bibinfo {author} {\bibfnamefont {P.-X.}\ \bibnamefont
  {Ma}}, \bibinfo {author} {\bibfnamefont {X.}~\bibnamefont {Feng}}, \bibinfo
  {author} {\bibfnamefont {M.}~\bibnamefont {Gorchtein}}, \bibinfo {author}
  {\bibfnamefont {L.-C.}\ \bibnamefont {Jin}},\ and\ \bibinfo {author}
  {\bibfnamefont {C.-Y.}\ \bibnamefont {Seng}},\ }\bibfield  {title} {\bibinfo
  {title} {{Lattice QCD calculation of the electroweak box diagrams for the
  kaon semileptonic decays}},\ }\href
  {https://doi.org/10.1103/PhysRevD.103.114503} {\bibfield  {journal} {\bibinfo
   {journal} {Phys.\ Rev.}\ }\textbf {\bibinfo {volume} {D103}},\ \bibinfo
  {pages} {114503} (\bibinfo {year} {2021})},\ \Eprint
  {https://arxiv.org/abs/2102.12048} {arXiv:2102.12048 [hep-lat]} \BibitemShut
  {NoStop}%
\bibitem [{\citenamefont {Czarnecki}\ \emph {et~al.}(2018)\citenamefont
  {Czarnecki}, \citenamefont {Marciano},\ and\ \citenamefont
  {Sirlin}}]{Czarnecki:2018okw}%
  \BibitemOpen
  \bibfield  {author} {\bibinfo {author} {\bibfnamefont {A.}~\bibnamefont
  {Czarnecki}}, \bibinfo {author} {\bibfnamefont {W.~J.}\ \bibnamefont
  {Marciano}},\ and\ \bibinfo {author} {\bibfnamefont {A.}~\bibnamefont
  {Sirlin}},\ }\bibfield  {title} {\bibinfo {title} {Neutron lifetime and axial
  coupling connection},\ }\href
  {https://doi.org/10.1103/PhysRevLett.120.202002} {\bibfield  {journal}
  {\bibinfo  {journal} {Phys.\ Rev.\ Lett.}\ }\textbf {\bibinfo {volume}
  {120}},\ \bibinfo {pages} {202002} (\bibinfo {year} {2018})}\BibitemShut
  {NoStop}%
\bibitem [{\citenamefont {Cirigliano}\ \emph {et~al.}(2003)\citenamefont
  {Cirigliano}, \citenamefont {Knecht}, \citenamefont {Neufeld},\ and\
  \citenamefont {Pichl}}]{Cirigliano:2002ng}%
  \BibitemOpen
  \bibfield  {author} {\bibinfo {author} {\bibfnamefont {V.}~\bibnamefont
  {Cirigliano}}, \bibinfo {author} {\bibfnamefont {M.}~\bibnamefont {Knecht}},
  \bibinfo {author} {\bibfnamefont {H.}~\bibnamefont {Neufeld}},\ and\ \bibinfo
  {author} {\bibfnamefont {H.}~\bibnamefont {Pichl}},\ }\bibfield  {title}
  {\bibinfo {title} {{The Pionic beta decay in chiral perturbation theory}},\
  }\href {https://doi.org/10.1140/epjc/s2002-01093-2} {\bibfield  {journal}
  {\bibinfo  {journal} {Eur. Phys. J. C}\ }\textbf {\bibinfo {volume} {27}},\
  \bibinfo {pages} {255} (\bibinfo {year} {2003})},\ \Eprint
  {https://arxiv.org/abs/hep-ph/0209226} {arXiv:hep-ph/0209226} \BibitemShut
  {NoStop}%
\bibitem [{\citenamefont {Pocanic}\ \emph {et~al.}(2004)\citenamefont {Pocanic}
  \emph {et~al.}}]{Pocanic:2003pf}%
  \BibitemOpen
  \bibfield  {author} {\bibinfo {author} {\bibfnamefont {D.}~\bibnamefont
  {Pocanic}} \emph {et~al.},\ }\bibfield  {title} {\bibinfo {title} {{Precise
  measurement of the pi+ ---\ensuremath{>} pi0 e+ nu branching ratio}},\ }\href
  {https://doi.org/10.1103/PhysRevLett.93.181803} {\bibfield  {journal}
  {\bibinfo  {journal} {Phys. Rev. Lett.}\ }\textbf {\bibinfo {volume} {93}},\
  \bibinfo {pages} {181803} (\bibinfo {year} {2004})},\ \Eprint
  {https://arxiv.org/abs/hep-ex/0312030} {arXiv:hep-ex/0312030} \BibitemShut
  {NoStop}%
\bibitem [{\citenamefont {Czarnecki}\ \emph {et~al.}(2020)\citenamefont
  {Czarnecki}, \citenamefont {Marciano},\ and\ \citenamefont
  {Sirlin}}]{Czarnecki:2019iwz}%
  \BibitemOpen
  \bibfield  {author} {\bibinfo {author} {\bibfnamefont {A.}~\bibnamefont
  {Czarnecki}}, \bibinfo {author} {\bibfnamefont {W.~J.}\ \bibnamefont
  {Marciano}},\ and\ \bibinfo {author} {\bibfnamefont {A.}~\bibnamefont
  {Sirlin}},\ }\bibfield  {title} {\bibinfo {title} {{Pion beta decay and
  Cabibbo-Kobayashi-Maskawa unitarity}},\ }\href
  {https://doi.org/10.1103/PhysRevD.101.091301} {\bibfield  {journal} {\bibinfo
   {journal} {Phys.\ Rev.}\ }\textbf {\bibinfo {volume} {D101}},\ \bibinfo
  {pages} {091301} (\bibinfo {year} {2020})},\ \Eprint
  {https://arxiv.org/abs/1911.04685} {arXiv:1911.04685 [hep-ph]} \BibitemShut
  {NoStop}%
\bibitem [{\citenamefont {Altmannshofer}\ \emph {et~al.}(2022)\citenamefont
  {Altmannshofer} \emph {et~al.}}]{PIONEER:2022yag}%
  \BibitemOpen
  \bibfield  {author} {\bibinfo {author} {\bibfnamefont {W.}~\bibnamefont
  {Altmannshofer}} \emph {et~al.} (\bibinfo {collaboration} {PIONEER}),\
  }\href@noop {} {\bibinfo {title} {{PIONEER: Studies of Rare Pion Decays}}}
  (\bibinfo {year} {2022}),\ \Eprint {https://arxiv.org/abs/2203.01981}
  {arXiv:2203.01981 [hep-ex]} \BibitemShut {NoStop}%
\bibitem [{\citenamefont {Aoki}\ \emph
  {et~al.}(2022{\natexlab{b}})\citenamefont {Aoki} \emph
  {et~al.}}]{FlavourLatticeAveragingGroupFLAG:2021npn}%
  \BibitemOpen
  \bibfield  {author} {\bibinfo {author} {\bibfnamefont {Y.}~\bibnamefont
  {Aoki}} \emph {et~al.} (\bibinfo {collaboration} {Flavour Lattice Averaging
  Group (FLAG)}),\ }\bibfield  {title} {\bibinfo {title} {{FLAG Review 2021}},\
  }\href {https://doi.org/10.1140/epjc/s10052-022-10536-1} {\bibfield
  {journal} {\bibinfo  {journal} {Eur.\ Phys.\ J.}\ }\textbf {\bibinfo {volume}
  {C82}},\ \bibinfo {pages} {869} (\bibinfo {year} {2022}{\natexlab{b}})},\
  \Eprint {https://arxiv.org/abs/2111.09849} {arXiv:2111.09849 [hep-lat]}
  \BibitemShut {NoStop}%
\bibitem [{\citenamefont {Seng}\ \emph {et~al.}(2022)\citenamefont {Seng},
  \citenamefont {Galviz}, \citenamefont {Marciano},\ and\ \citenamefont
  {Mei\ss{}ner}}]{Seng:2021nar}%
  \BibitemOpen
  \bibfield  {author} {\bibinfo {author} {\bibfnamefont {C.-Y.}\ \bibnamefont
  {Seng}}, \bibinfo {author} {\bibfnamefont {D.}~\bibnamefont {Galviz}},
  \bibinfo {author} {\bibfnamefont {W.~J.}\ \bibnamefont {Marciano}},\ and\
  \bibinfo {author} {\bibfnamefont {U.-G.}\ \bibnamefont {Mei\ss{}ner}},\
  }\bibfield  {title} {\bibinfo {title} {{Update on |Vus| and |Vus/Vud| from
  semileptonic kaon and pion decays}},\ }\href
  {https://doi.org/10.1103/PhysRevD.105.013005} {\bibfield  {journal} {\bibinfo
   {journal} {Phys.\ Rev.}\ }\textbf {\bibinfo {volume} {D105}},\ \bibinfo
  {pages} {013005} (\bibinfo {year} {2022})},\ \Eprint
  {https://arxiv.org/abs/2107.14708} {arXiv:2107.14708 [hep-ph]} \BibitemShut
  {NoStop}%
\bibitem [{\citenamefont {Seng}\ \emph
  {et~al.}(2019{\natexlab{b}})\citenamefont {Seng}, \citenamefont {Gorchtein},\
  and\ \citenamefont {Ramsey-Musolf}}]{PhysRevD.100.013001}%
  \BibitemOpen
  \bibfield  {author} {\bibinfo {author} {\bibfnamefont {C.-Y.}\ \bibnamefont
  {Seng}}, \bibinfo {author} {\bibfnamefont {M.}~\bibnamefont {Gorchtein}},\
  and\ \bibinfo {author} {\bibfnamefont {M.~J.}\ \bibnamefont
  {Ramsey-Musolf}},\ }\bibfield  {title} {\bibinfo {title} {Dispersive
  evaluation of the inner radiative correction in neutron and nuclear
  $\ensuremath{\beta}$ decay},\ }\href
  {https://doi.org/10.1103/PhysRevD.100.013001} {\bibfield  {journal} {\bibinfo
   {journal} {Phys. Rev. D}\ }\textbf {\bibinfo {volume} {100}},\ \bibinfo
  {pages} {013001} (\bibinfo {year} {2019}{\natexlab{b}})}\BibitemShut
  {NoStop}%
\bibitem [{\citenamefont {Gupta}\ \emph {et~al.}(2018)\citenamefont {Gupta},
  \citenamefont {Jang}, \citenamefont {Yoon}, \citenamefont {Lin},
  \citenamefont {Cirigliano},\ and\ \citenamefont
  {Bhattacharya}}]{Gupta:2018qil}%
  \BibitemOpen
  \bibfield  {author} {\bibinfo {author} {\bibfnamefont {R.}~\bibnamefont
  {Gupta}}, \bibinfo {author} {\bibfnamefont {Y.-C.}\ \bibnamefont {Jang}},
  \bibinfo {author} {\bibfnamefont {B.}~\bibnamefont {Yoon}}, \bibinfo {author}
  {\bibfnamefont {H.-W.}\ \bibnamefont {Lin}}, \bibinfo {author} {\bibfnamefont
  {V.}~\bibnamefont {Cirigliano}},\ and\ \bibinfo {author} {\bibfnamefont
  {T.}~\bibnamefont {Bhattacharya}},\ }\bibfield  {title} {\bibinfo {title}
  {{Isovector Charges of the Nucleon from 2+1+1-flavor Lattice QCD}},\ }\href
  {https://doi.org/10.1103/PhysRevD.98.034503} {\bibfield  {journal} {\bibinfo
  {journal} {Phys.\ Rev.}\ }\textbf {\bibinfo {volume} {D98}},\ \bibinfo
  {pages} {034503} (\bibinfo {year} {2018})},\ \Eprint
  {https://arxiv.org/abs/1806.09006} {arXiv:1806.09006 [hep-lat]} \BibitemShut
  {NoStop}%
\bibitem [{\citenamefont {Bazavov}\ \emph {et~al.}(2013)\citenamefont {Bazavov}
  \emph {et~al.}}]{Bazavov:2012xda}%
  \BibitemOpen
  \bibfield  {author} {\bibinfo {author} {\bibfnamefont {A.}~\bibnamefont
  {Bazavov}} \emph {et~al.} (\bibinfo {collaboration} {MILC Collaboration}),\
  }\bibfield  {title} {\bibinfo {title} {{Lattice QCD ensembles with four
  flavors of highly improved staggered quarks}},\ }\href
  {https://doi.org/10.1103/PhysRevD.87.054505} {\bibfield  {journal} {\bibinfo
  {journal} {Phys.\ Rev.}\ }\textbf {\bibinfo {volume} {D87}},\ \bibinfo
  {pages} {054505} (\bibinfo {year} {2013})},\ \Eprint
  {https://arxiv.org/abs/1212.4768} {arXiv:1212.4768 [hep-lat]} \BibitemShut
  {NoStop}%
\bibitem [{\citenamefont {Deur}\ \emph {et~al.}(2016)\citenamefont {Deur},
  \citenamefont {Brodsky},\ and\ \citenamefont {de~Teramond}}]{Deur:2016tte}%
  \BibitemOpen
  \bibfield  {author} {\bibinfo {author} {\bibfnamefont {A.}~\bibnamefont
  {Deur}}, \bibinfo {author} {\bibfnamefont {S.~J.}\ \bibnamefont {Brodsky}},\
  and\ \bibinfo {author} {\bibfnamefont {G.~F.}\ \bibnamefont {de~Teramond}},\
  }\bibfield  {title} {\bibinfo {title} {{The QCD Running Coupling}},\ }\href
  {https://doi.org/10.1016/j.ppnp.2016.04.003} {\bibfield  {journal} {\bibinfo
  {journal} {Nucl.\ Phys.}\ }\textbf {\bibinfo {volume} {B90}},\ \bibinfo
  {pages} {1} (\bibinfo {year} {2016})},\ \Eprint
  {https://arxiv.org/abs/1604.08082} {arXiv:1604.08082 [hep-ph]} \BibitemShut
  {NoStop}%
\bibitem [{\citenamefont {Patrignani}\ \emph {et~al.}(2016)\citenamefont
  {Patrignani} \emph {et~al.}}]{ParticleDataGroup:2016lqr}%
  \BibitemOpen
  \bibfield  {author} {\bibinfo {author} {\bibfnamefont {C.}~\bibnamefont
  {Patrignani}} \emph {et~al.} (\bibinfo {collaboration} {Particle Data
  Group}),\ }\bibfield  {title} {\bibinfo {title} {{Review of Particle
  Physics}},\ }\href {https://doi.org/10.1088/1674-1137/40/10/100001}
  {\bibfield  {journal} {\bibinfo  {journal} {Chin. Phys. C}\ }\textbf
  {\bibinfo {volume} {40}},\ \bibinfo {pages} {100001} (\bibinfo {year}
  {2016})}\BibitemShut {NoStop}%
\bibitem [{\citenamefont {Park}\ \emph {et~al.}(2022)\citenamefont {Park},
  \citenamefont {Gupta}, \citenamefont {Yoon}, \citenamefont {Mondal},
  \citenamefont {Bhattacharya}, \citenamefont {Jang}, \citenamefont {Jo\'o},\
  and\ \citenamefont {Winter}}]{Park:2021ypf}%
  \BibitemOpen
  \bibfield  {author} {\bibinfo {author} {\bibfnamefont {S.}~\bibnamefont
  {Park}}, \bibinfo {author} {\bibfnamefont {R.}~\bibnamefont {Gupta}},
  \bibinfo {author} {\bibfnamefont {B.}~\bibnamefont {Yoon}}, \bibinfo {author}
  {\bibfnamefont {S.}~\bibnamefont {Mondal}}, \bibinfo {author} {\bibfnamefont
  {T.}~\bibnamefont {Bhattacharya}}, \bibinfo {author} {\bibfnamefont {Y.-C.}\
  \bibnamefont {Jang}}, \bibinfo {author} {\bibfnamefont {B.}~\bibnamefont
  {Jo\'o}},\ and\ \bibinfo {author} {\bibfnamefont {F.}~\bibnamefont {Winter}}
  (\bibinfo {collaboration} {Nucleon Matrix Elements (NME)}),\ }\bibfield
  {title} {\bibinfo {title} {{Precision nucleon charges and form factors using
  (2+1)-flavor lattice QCD}},\ }\href
  {https://doi.org/10.1103/PhysRevD.105.054505} {\bibfield  {journal} {\bibinfo
   {journal} {Phys. Rev. D}\ }\textbf {\bibinfo {volume} {105}},\ \bibinfo
  {pages} {054505} (\bibinfo {year} {2022})},\ \Eprint
  {https://arxiv.org/abs/2103.05599} {arXiv:2103.05599 [hep-lat]} \BibitemShut
  {NoStop}%
\bibitem [{\citenamefont {Virtanen}\ \emph {et~al.}(2020)\citenamefont
  {Virtanen}, \citenamefont {Gommers}, \citenamefont {Oliphant}, \citenamefont
  {Haberland}, \citenamefont {Reddy}, \citenamefont {Cournapeau}, \citenamefont
  {Burovski}, \citenamefont {Peterson}, \citenamefont {Weckesser},
  \citenamefont {Bright}, \citenamefont {{van der Walt}}, \citenamefont
  {Brett}, \citenamefont {Wilson}, \citenamefont {Millman}, \citenamefont
  {Mayorov}, \citenamefont {Nelson}, \citenamefont {Jones}, \citenamefont
  {Kern}, \citenamefont {Larson}, \citenamefont {Carey}, \citenamefont {Polat},
  \citenamefont {Feng}, \citenamefont {Moore}, \citenamefont {{VanderPlas}},
  \citenamefont {Laxalde}, \citenamefont {Perktold}, \citenamefont {Cimrman},
  \citenamefont {Henriksen}, \citenamefont {Quintero}, \citenamefont {Harris},
  \citenamefont {Archibald}, \citenamefont {Ribeiro}, \citenamefont
  {Pedregosa}, \citenamefont {{van Mulbregt}},\ and\ \citenamefont {{SciPy 1.0
  Contributors}}}]{2020SciPy-NMeth}%
  \BibitemOpen
  \bibfield  {author} {\bibinfo {author} {\bibfnamefont {P.}~\bibnamefont
  {Virtanen}}, \bibinfo {author} {\bibfnamefont {R.}~\bibnamefont {Gommers}},
  \bibinfo {author} {\bibfnamefont {T.~E.}\ \bibnamefont {Oliphant}}, \bibinfo
  {author} {\bibfnamefont {M.}~\bibnamefont {Haberland}}, \bibinfo {author}
  {\bibfnamefont {T.}~\bibnamefont {Reddy}}, \bibinfo {author} {\bibfnamefont
  {D.}~\bibnamefont {Cournapeau}}, \bibinfo {author} {\bibfnamefont
  {E.}~\bibnamefont {Burovski}}, \bibinfo {author} {\bibfnamefont
  {P.}~\bibnamefont {Peterson}}, \bibinfo {author} {\bibfnamefont
  {W.}~\bibnamefont {Weckesser}}, \bibinfo {author} {\bibfnamefont
  {J.}~\bibnamefont {Bright}}, \bibinfo {author} {\bibfnamefont {S.~J.}\
  \bibnamefont {{van der Walt}}}, \bibinfo {author} {\bibfnamefont
  {M.}~\bibnamefont {Brett}}, \bibinfo {author} {\bibfnamefont
  {J.}~\bibnamefont {Wilson}}, \bibinfo {author} {\bibfnamefont {K.~J.}\
  \bibnamefont {Millman}}, \bibinfo {author} {\bibfnamefont {N.}~\bibnamefont
  {Mayorov}}, \bibinfo {author} {\bibfnamefont {A.~R.~J.}\ \bibnamefont
  {Nelson}}, \bibinfo {author} {\bibfnamefont {E.}~\bibnamefont {Jones}},
  \bibinfo {author} {\bibfnamefont {R.}~\bibnamefont {Kern}}, \bibinfo {author}
  {\bibfnamefont {E.}~\bibnamefont {Larson}}, \bibinfo {author} {\bibfnamefont
  {C.~J.}\ \bibnamefont {Carey}}, \bibinfo {author} {\bibfnamefont
  {{\.I}.}~\bibnamefont {Polat}}, \bibinfo {author} {\bibfnamefont
  {Y.}~\bibnamefont {Feng}}, \bibinfo {author} {\bibfnamefont {E.~W.}\
  \bibnamefont {Moore}}, \bibinfo {author} {\bibfnamefont {J.}~\bibnamefont
  {{VanderPlas}}}, \bibinfo {author} {\bibfnamefont {D.}~\bibnamefont
  {Laxalde}}, \bibinfo {author} {\bibfnamefont {J.}~\bibnamefont {Perktold}},
  \bibinfo {author} {\bibfnamefont {R.}~\bibnamefont {Cimrman}}, \bibinfo
  {author} {\bibfnamefont {I.}~\bibnamefont {Henriksen}}, \bibinfo {author}
  {\bibfnamefont {E.~A.}\ \bibnamefont {Quintero}}, \bibinfo {author}
  {\bibfnamefont {C.~R.}\ \bibnamefont {Harris}}, \bibinfo {author}
  {\bibfnamefont {A.~M.}\ \bibnamefont {Archibald}}, \bibinfo {author}
  {\bibfnamefont {A.~H.}\ \bibnamefont {Ribeiro}}, \bibinfo {author}
  {\bibfnamefont {F.}~\bibnamefont {Pedregosa}}, \bibinfo {author}
  {\bibfnamefont {P.}~\bibnamefont {{van Mulbregt}}},\ and\ \bibinfo {author}
  {\bibnamefont {{SciPy 1.0 Contributors}}},\ }\bibfield  {title} {\bibinfo
  {title} {{{SciPy} 1.0: Fundamental Algorithms for Scientific Computing in
  Python}},\ }\href {https://doi.org/10.1038/s41592-019-0686-2} {\bibfield
  {journal} {\bibinfo  {journal} {Nature Methods}\ }\textbf {\bibinfo {volume}
  {17}},\ \bibinfo {pages} {261} (\bibinfo {year} {2020})}\BibitemShut
  {NoStop}%
\bibitem [{\citenamefont {Larin}\ and\ \citenamefont
  {Vermaseren}(1991)}]{Larin:1991tj}%
  \BibitemOpen
  \bibfield  {author} {\bibinfo {author} {\bibfnamefont {S.~A.}\ \bibnamefont
  {Larin}}\ and\ \bibinfo {author} {\bibfnamefont {J.~A.~M.}\ \bibnamefont
  {Vermaseren}},\ }\bibfield  {title} {\bibinfo {title} {{The alpha-s**3
  corrections to the Bjorken sum rule for polarized electroproduction and to
  the Gross-Llewellyn Smith sum rule}},\ }\href
  {https://doi.org/10.1016/0370-2693(91)90839-I} {\bibfield  {journal}
  {\bibinfo  {journal} {Phys. Lett. B}\ }\textbf {\bibinfo {volume} {259}},\
  \bibinfo {pages} {345} (\bibinfo {year} {1991})}\BibitemShut {NoStop}%
\bibitem [{\citenamefont {Baikov}\ \emph {et~al.}(2010)\citenamefont {Baikov},
  \citenamefont {Chetyrkin},\ and\ \citenamefont {Kuhn}}]{Baikov:2010je}%
  \BibitemOpen
  \bibfield  {author} {\bibinfo {author} {\bibfnamefont {P.~A.}\ \bibnamefont
  {Baikov}}, \bibinfo {author} {\bibfnamefont {K.~G.}\ \bibnamefont
  {Chetyrkin}},\ and\ \bibinfo {author} {\bibfnamefont {J.~H.}\ \bibnamefont
  {Kuhn}},\ }\bibfield  {title} {\bibinfo {title} {{Adler Function, Bjorken Sum
  Rule, and the Crewther Relation to Order $\alpha^4_s$ in a General Gauge
  Theory}},\ }\href {https://doi.org/10.1103/PhysRevLett.104.132004} {\bibfield
   {journal} {\bibinfo  {journal} {Phys. Rev. Lett.}\ }\textbf {\bibinfo
  {volume} {104}},\ \bibinfo {pages} {132004} (\bibinfo {year} {2010})},\
  \Eprint {https://arxiv.org/abs/1001.3606} {arXiv:1001.3606 [hep-ph]}
  \BibitemShut {NoStop}%
\bibitem [{\citenamefont {Edwards}\ and\ \citenamefont
  {Joo}(2005)}]{Edwards:2004sx}%
  \BibitemOpen
  \bibfield  {author} {\bibinfo {author} {\bibfnamefont {R.~G.}\ \bibnamefont
  {Edwards}}\ and\ \bibinfo {author} {\bibfnamefont {B.}~\bibnamefont {Joo}}
  (\bibinfo {collaboration} {SciDAC Collaboration, LHPC Collaboration, UKQCD
  Collaboration}),\ }\bibfield  {title} {\bibinfo {title} {{The Chroma software
  system for lattice QCD}},\ }\href
  {https://doi.org/10.1016/j.nuclphysbps.2004.11.254} {\bibfield  {journal}
  {\bibinfo  {journal} {Nucl.\ Phys.\ Proc.\ Suppl.}\ }\textbf {\bibinfo
  {volume} {140}},\ \bibinfo {pages} {832} (\bibinfo {year} {2005})},\ \Eprint
  {https://arxiv.org/abs/hep-lat/0409003} {arXiv:hep-lat/0409003 [hep-lat]}
  \BibitemShut {NoStop}%
\end{thebibliography}%

\end{document}